\documentclass[%
 reprint,
 amsmath,amssymb,
 aps,
]{revtex4-1}
\usepackage{graphicx,epstopdf}% Include figure files
\usepackage{dcolumn}% Align table columns on decimal point
\usepackage{bm}% bold math
\usepackage{hyperref}% add hypertext capabilities
\usepackage{url}
\newcommand{\average}[1]{\ensuremath{\langle#1\rangle} }

\begin{document}

\preprint{APS/123-QED}

\title{Magnetic correlation effects by the topological zero mode \\
in a hydrogenated graphene vacancy $V_{111}$}% Force line breaks with \\

\author{Naoki Morishita}
 \email{morishi@artemis.mp.es.osaka-u.ac.jp}
\author{Gagus Ketut Sunnardianto}
 \email{gagus@artemis.mp.es.osaka-u.ac.jp}
\author{Koichi Kusakabe}%
 \email{kabe@mp.es.osaka-u.ac.jp}
\affiliation{%
Graduate School of Engineering Science, Osaka University,\\
1-3 Machikaneyama-cho, Toyonaka, Osaka , 560-8531 Japan.
}%

\author{Isao Maruyama}
\affiliation{
Faculty of Information Engineering, Fukuoka Institute of Technology,\\
 3-30-1Wajiro, Higashi, Higashi-ku, Fukuoka, 811-0295,Japan.
}%
\author{Kazuyuki Takai}
\affiliation{%
Faculty of Bioscience and Applied Chemistry, Hosei University,\\
3-7-2 Kajinocho, Koganei, Tokyo, 184-8584, Japan.
}%
\author{Toshiaki Enoki}
\affiliation{%
Department of Chemistry, Tokyo Institute of Technology,\\
2-12-1 Ookayama, Meguro-ku, Tokyo, 152-8551, Japan.
}%

\date{\today}% It is always \today, today,

\begin{abstract}
Electron correlation effects caused by the topological zero mode 
of a hydrogenated graphene vacancy, $V_{111}$, 
with three adsorbed hydrogen atoms is discussed theoretically. 
A Kondo model is derived from the multi-reference representation 
of the density functional theory, where exchange scattering processes 
between the zero mode and low-energy modes in the Dirac cones are estimated. 
Even when the Dirac cone is slightly off from the charge neutral point, 
a finite on-site correlation energy, $U_0$, 
for the zero mode of an isolated $V_{111}$ allows the half-filling of the localized 
level giving a spin $s=1/2$.  The anti-ferromagnetic Kondo screening 
mediated by higher order scattering processes becomes dominant 
in the dilute limit of the vacancies. 
Our estimation of relevant two body interactions 
certifies appearance of the Kondo effect at low temperatures. 
\end{abstract}

\pacs{Valid PACS appear here}% PACS, the Physics and Astronomy
\maketitle

%\tableofcontents

%%%%%%%%%%%%%%%%%%%%%%%%%%%%%%%%%%%%%%%%%%%%%%%%%%%%%%
\section{\label{sec:level1}Introduction}

After the finding of graphene and its unique properties,\cite{Novoselov-Science,Novoselov-Nature,RevModPhys.81.109} 
there has been a lot of discussions on possible electronic correlation effects in various defects of graphene. 
In theory, correlation effects in defect-induced magnetism were treated by applying 
several types of models.\cite{PhysRevLett.92.225502,PhysRevLett.96.036801,PhysRevB.83.241408,0034-4885-76-3-032501} 
Actually, the Kondo effect in graphene was highlighted for magnetic adatoms.\cite{PhysRevB.76.115407,PhysRevB.77.045417,PhysRevB.82.085423,PhysRevLett.101.026805,PhysRevLett.102.046801,
PhysRevLett.103.206804,PhysRevLett.106.016801,PhysRevB.85.115405,1367-2630-15-5-053018}
Furthermore, vacancy induced magnetism was focused after experimental findings were annouced,\cite{PhysRevLett.104.096804,PhysRevB.82.153414,
Nair-Sepioni-Tsai-Lehtinen-Keinonen-Krasheninnikov-Thomson-Geim-Grigorieva,PhysRevLett.109.186604,Chen2011} 
for which the Kondo effect was discussed theoretically.\cite{PhysRevB.83.241408,doi:10.1143/JPSJ.81.063709,2012arXiv1207.3135C,PhysRevB.88.075104} 
Theoretical analysis of possible defective graphene structures has been continuously performed, 
where hydrogenation of an atomic vacancy was analyzed by 
theoretical simulations.\cite{PhysRevLett.93.187202,PhysRevB.75.125408,PhysRevB.85.245443,PhysRevLett.104.096804}

Recently, identification of hydrogenated graphene vacancies 
(HGV) was successfully performed utilizing the high-resolution 
scanning tunneling microscopy 
supported by electronic structure calculations based on the density 
functional theory.\cite{PhysRevB.89.155405,PhysRevB.87.115427} 
This experimental result revives researchers' attention to HGV, 
among various defect structures of 
graphene.\cite{RevModPhys.81.109,0034-4885-76-3-032501} 
Atomic hydrogen treatment after Ar$^+$ irradiation on 
a crystalline graphitic sample is indeed an important method to create 
various types of defects including hydrogenated atomic vacancy 
and nano-holes in a controlled manner. 

In the experiment,\cite{PhysRevB.89.155405} 
two stable HGV structures, called $V_{111}$ and $V_{211}$, were identified. 
The former, which is a triply hydrogenated 
atomic vacancy, possesses a topological localized state, 
{\it i.e.} the zero mode, according to the local topological network of 
the $\pi$ electron system. While, the latter quadruply hydrogenated 
atomic vacancy does not have the zero mode. 
Therefore, the former structure of $V_{111}$ may have a localized spin $s=1/2$. 
Furthermore, electron correlation effects including the Kondo effect 
may be expected, similar to the bare vacancy. 

The zero-energy non-bonding localized state 
appears owing to topological characteristic of the $\pi$ network around 
$V_{111}$. Namely, around the vacancy, the number of the $B$ sublattice, 
$N_B$, is locally less than that of the $A$ sublattice, $N_A$, 
owing to the vacancy, assuming that the original graphene 
has balanced sublattices. In the argument of network topology of the $\pi$ system, 
the difference of $N_A-N_B=1$ certifies appearance of 
the $\pi$-non-bonding state at the Dirac point. 
These features of the special localized electron mode 
are completely reproduced in the band structure calculations. 
The simulated STM image is almost perfectly identical to the real observation, 
where usage of the deformed atomic structure 
optimized by the simulation is essential. 
Thus, the zero mode localized around the $V_{111}$ structure is 
confirmed both in experiment and by theory.\cite{PhysRevB.89.155405}  

The location of the zero mode of $V_{111}$ along the energy axis 
is at the Dirac point. This is always the case, when we simulate 
a neutral graphene with the triply hydrogenated atomic vacancy. 
In this condition, the occupation of the zero mode is just at the half-filling. 
Therefore, a magnetic moment of $s=1/2$ may appear in the real system, 
as a Kramers doublet. 
The correlation effect should be relevant for the magnetic property of $V_{111}$, 
which may give an isolated magnetic impurity. 
To confirm this picture, therefore, we need to go one step further 
beyond a single-particle description of 
the ordinary density functional theory (DFT). 

In this communication, we propose a method to model 
the correlated electron system of $V_{111}$. 
By evaluating the single-particle spectrum and two-particle 
effective scattering amplitudes, 
we derive a kind of the $s$-$d$ exchange model\cite{PhysRev.81.440} 
or a Kondo Hamiltonian\cite{Kondo} describing the action of 
the zero mode appearing in the hydrogenated graphene vacancy. 
The Kohn-Sham orbitals obtained by the local density 
approximation\cite{PhysRev.140.A1133} 
are used to estimate the scattering amplitudes. 
Here, relevant two-body interactions include the on-site repulsive 
interaction on the zero mode and effective exchange scatterings 
between the zero mode and the Dirac electrons. 

Converting the exchange operator into a separable form, 
we obtain an effective Anderson model with two local orbitals. 
Applying the continuous-time quantum Monte-Carlo calculation, 
we estimate the chemical potential dependence of the electron number 
on these local orbitals. 
The result suggests that 
the Kondo screening of the spin 1/2 at the zero mode can happen. 
In our representation, low-lying orbitals in the Dirac cones 
are correlated with the zero mode, forming a local singlet pair 
in the thermodynamic limit. This is a result of magnetic screening 
mediated by a kind of anti-ferromagnetic super exchange. 
However, the total ground state remains to be a doublet, which 
appears as a half-filled renormalized spectrum of the Dirac cones.  
The state should keep the zero-gap semiconducting feature of the graphene. 

%%%%%%%%%%%%%%%%%%%%%%%%%%%%%%%%%%%%%%%%%%%%%%%%%%%%%%%%

%%%%%%%%%%%%%%%%%%%%%%%%%%%%%%%%%%%%%%%%%%%%%%%%%%%%%%%%
\section{\label{sec:level1_2}Models of the zero mode on $V_{111}$}

\subsection{\label{sec:zero_mode_V111}The zero mode}

%%% The pi network of V111

At the triply hydrogenated vacancy, 
each hydrogen atom of $V_{111}$ forms 
a $\sigma$ bond with a carbon atom. 
We have perfect termination of carbon 
$\sigma$ dangling bonds at this vacancy. 
All of the carbon atoms have saturated $\sigma$ bonds, 
and there is a $\pi$ orbital at each carbon atom.  

\begin{figure}[htbp] 
\begin{center} 
\includegraphics[width=6cm]{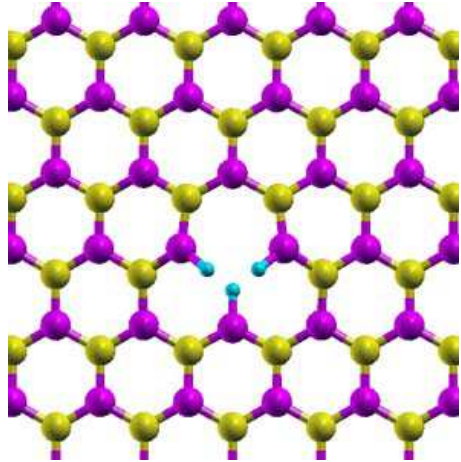} 
\end{center} 
\caption{The atomic structure of $V_{111}$. Carbon atoms are grouped into 
subsets, $A$ and $B$, where atoms in $A$ and $B$ are colored 
magenta and yellow, respectively. Hydrogen atoms are in cyan. 
At each carbon site, we have a $\pi$ orbital. The connections, each of which 
links two neighboring carbon atoms in a pair, give 
a defective honeycomb network of $V_{111}$. 
This drawing is created by using XCrySDen.\cite{Kokalj2003155}} 
\label{fig:Fig.0} 
\end{figure}

Here, lobes of the $\pi$ orbital is oriented in a direction 
nearly perpendicular to the local $sp^2$ $\sigma$ bonds. 
To describe a basic idea on 
the $\pi$ topology, at first, we neglect the deformation of 
graphene sheet around $V_{111}$. 
By drawing connections between neighboring $\pi$ orbitals, 
we have a honeycomb network with a vacancy. (Fig.~\ref{fig:Fig.0}) 

To avoid effects of graphene edges, we may use 
the periodic boundary condition along two in-plane directions. 
The structure contains $N_C$ carbon atoms and three hydrogen atoms. 
The carbon network gives a bipartite graph, where 
all vertices are grouped in one of two subsets, $A$ or $B$, 
such that two neighboring carbons connected by an edge 
are contained in the other subsets, {\it i.e.} one in $A$ and the other in $B$. 
Owing to the vacancy site, which is in the $B$ sublattice, 
the number of $A$ sublattice sites is 
larger than that of $B$ sublattice sites by 1, 
{\it i.e.} $N_A-N_B=1$. We have $N_C=N_A+N_B$. 

When we adopt a scheme of tight-binding description for a pure $\pi$ system, 
assuming that transfer parameters are nearly the same 
and are non-zero only for each neighboring pair of carbon sites, 
we have a tight-binding model, {\it i.e.} the $t$ only model. 
Introducing the creation (and annihilation) operators, 
$C_{i,\sigma}^\dagger$ ($C_{i,\sigma}$), of a $\pi$ electron at the $i$-th 
carbon site, we have a Hamiltonian operator of the $t$ model, 
\begin{eqnarray}
\hat{H}_{TBM} 
&=& - t \sum_{\langle i,j\rangle}
\sum_{\sigma=\uparrow,\downarrow} 
\left\{C_{i,\sigma}^\dagger C_{j,\sigma} + H.c.\right\} \nonumber \\
&=& \sum_{\sigma=\uparrow,\downarrow} 
\left( C_{i_A,\sigma}^\dagger , C_{j_B,\sigma}^\dagger \right)
H_{TBM}
\left( \begin{array}{c} C_{i_A,\sigma} \\ C_{j_B,\sigma} \end{array} \right) \nonumber \\
&=& \sum_{k=-N_B}^{N_B}\sum_{\sigma=\uparrow,\downarrow} 
\tilde{\varepsilon}_{k} c_{k,\sigma}^\dagger c_{k,\sigma}, \\
H_{TBM} &=&
\left( 
\begin{array}{cc} 0 & H_{AB} \\ H_{BA} & 0 \end{array}
\right) . \nonumber 
\end{eqnarray}
Here, a symbol of $\langle i,j\rangle$ is used to specify 
a summation with respect to the neighboring pair of carbon atoms. 
In the above expression of $\hat{H}_{TBM}$, 
the diagonal representation by $c_{k,\sigma}^\dagger$ 
and $c_{k,\sigma}$ is given by the unitary transformation 
to perform the diagonalization of $H_{TBM}$. 
The eigen energy, $\tilde{\varepsilon}_{k}$, gives the spectrum of 
the $t$ model on the honeycomb network with a vacancy. 

Here, the matrix $H_{TBM}$ is an $(N_C \times N_C)$ matrix 
with an $(N_A\times N_B)$ submatrix $H_{AB}$ representing 
hopping processes from a $B$ site to an $A$ site. 
The index $i_A$ (or $j_B$) runs only in the $A$ (or $B$) sublattice sites 
and $\left( C_{i_A,\sigma}^\dagger , C_{j_B,\sigma}^\dagger \right)$ 
is treated as a $(1\times N_C)$ vector form. 
The matrix $H_{TBM}$ has $(N_A\times N_A)$ and $(N_B\times N_B)$ 
zero matrices as its diagonal submatrices, 
which represents the bipartiteness leading us to 
a particle-hole symmetry. 
The quasi spinor representation allows us to have 
a gauge transformation, 
\[ C_{i_A,\sigma}^\dagger \longrightarrow  d_{i_A,\sigma}, 
\qquad C_{j_B,\sigma}^\dagger \longrightarrow  (-1) \times d_{j_B,\sigma}, \]
which gives 
\begin{eqnarray}
\label{TBM0}
\hat{H}_{TBM} 
&=& \sum_{\sigma=\uparrow,\downarrow} 
\left( d_{i_A,\sigma}^\dagger , d_{j_B,\sigma}^\dagger \right)
H_{TBM}
\left( \begin{array}{c} d_{i_A,\sigma} \\ d_{j_B,\sigma} \end{array} \right) 
\nonumber \\
&=& - t \sum_{\langle i,j\rangle}
\sum_{\sigma=\uparrow,\downarrow} 
\left\{d_{i,\sigma}^\dagger d_{j,\sigma} + H.c.\right\} , 
\end{eqnarray}
so that the spectrum of the hole Hamiltonian is the same as the original. 
With a single vacancy in the honeycomb network, $N_A-N_B=1$, 
since the rank of the matrix $H_{TBM}$ is no more than 
$2N_B=N_A+N_B-1=N_C-1$, 
a conclusion is that a zero eigen value appears 
at the center of the spectrum. 
This argument is well known.\cite{PhysRevLett.62.1201} 
Furthermore, when the spectrum of $H_{TBM}$ has a discrete nature, 
the zero eigen mode is stable against perturbation.
In its diagonal form, the spectrum of $\hat{H}_{TBM}$, $\tilde{\varepsilon}_{k}$, 
is specified by an integer, $-N_B\le k \le N_B$. 
IN the $t$ model, $\tilde{\varepsilon}_{-k}=-\tilde{\varepsilon}_k$ 
owing to the particle-hole symmetry. 
The $k=0$ mode is the zero mode with $\tilde{\varepsilon}_{0}=0$. 

This zero mode of the $t$ model was analyzed in its nature 
as a non-bonding damping mode.\cite{PhysRevLett.96.036801} 
In that study, the authors considered a $t$-$t'$ model 
with a one-dimensional boundary condition of a semi-infinite system. 
The spectrum they considered 
was continuous at the Dirac point, where 
the upper and lower continuum in the spectrum 
touched the zero mode peak in a local density of state. 
When the particle-hole symmetry is broken, 
the zero mode became a resonance mode. 

Let's consider a supercell structure of $N_C$ 
carbon atoms of $V_{111}$ in a unit cell. 
We assume that there are $N_{cell}$ supercells as a whole, 
adopting the Born-von-Karman periodic boundary condition. 
For the $t$ model, owing to the unbalanced sublattice structures, 
the total system has $N_{cell} \times (N_A-N_B)$ zero modes. 
These zero modes provide an exactly flat band for the $t$ model. 

Owing to a finite size of the cell, 
a gap at the Dirac point can be formed. 
As for the $t$ model, we may make a choice of the supercell 
in a rhomboid form with $N_C=2\times(3n\pm1)^2-1$. 
Here, $n$ is an integer. 
We will call the cell a $(3n\pm1)\times (3n\pm1)$ cell. 
On this system, the band structure shows a gap at the K and K' points 
of the folded Brillouin zone (FBZ). 
The flat band appears as ingap states. 
(For comparison with the DFT simulation, 
see also Appendix~\ref{Appendix_DFT_LDA}.) 
In this model $V_{111}$ structure, 
we expect stability of the flat band against small perturbation. 
As for the other $3n\times 3n$ supercell with $N_C=2\times(3n)^2-1$, 
the flat band touches the Dirac cones at the $\Gamma$ point. 
This is owing to an adjusted periodicity of Bloch wave functions 
at $K$ (or $K'$), which allows one of recombined Dirac-cone branches 
unaffected by the defect. 

Similarly, we might have a nearly flat band in a generalized 
tight-binding model of $V_{111}$, where long-range transfer terms, 
$s$ and $p$ orbitals are considered. 
Using a parameter sets of Papaconstantopoulos with 
additional parameters for C-H $\sigma$ 
bonds,\cite{Papaconstantopoulos,PhysRevB.71.193406} 
we have obtained band structures for $V_{111}$. 
In the actual $V_{111}$ structure, we have slight corrugation of 
graphene plane around the vacancy. 
This is due to the geometric exclusion among hydrogen atoms 
coming from the inter-atomic interaction. 
Therefore, for the calculation of the $s$-$p$ model, 
we utilize an atomic structure obtained by an optimization simulation 
in a DFT-Kohn-Sham local-density-approximation (LDA) 
scheme.\cite{PhysRev.136.B864,PhysRev.140.A1133} 
The obtained band structures are roughly similar to that of the $t$ model. 
When we see supercells of $n=3, 4$, 
we have always a nearly flat band at the Fermi level, 

The similarity in the band structures suggests that we have 
a mapping from an eigen state of the $s$-$p$ model to 
a counterpart of the $t$ model, 
at least when the eigen modes around the Fermi level are concerned. 
The $(N_C+1)/2+(3N_C+3)/2$-th band of 
the $s$-$p$ model may be a correspondence to 
the $(N_C+1)/2$-th band of the $t$ model. 
Here, for a definition of the mapping, 
a $\pi$ orbital at each carbon site is required. 
Actually, it is possible to define a local $\pi$ orbital in 
the deformed graphene structure of $V_{111}$. 
(See a method in Appendix~\ref{Appendix_pi_orbital}.) 
Then, our whole simulations 
suggest that there is a well-defined topological mapping 
between an $s$-$p$ model and the $t$ model as far as 
$V_{111}$ in a finite cell structure with 
the periodic boundary condition is considered. 

The mapping among low-lying phase spaces of different effective models 
allows us to call the low-lying flat band ``the zero modes''. 
When we take the limit of large supercell size, or $N_C \rightarrow \infty$, 
the true relevance of the idea will appear. 
Before going to this point, let us introduce the DFT simulation 
and another effective many-body Hamiltonian. 

\subsection{\label{sec:level2_2-0}Scaling of effective Hamiltonians}

All of reported band structure calculations of $V_{111}$ 
in a super cell, based on the density functional theory, 
told that a flat band appears around the Dirac 
point.\cite{PhysRevB.85.245443,PhysRevB.89.155405}  
In this band, each eigen state of the Kohn-Sham Hamiltonian 
has a non-bonding character, where the wave function 
has a finite amplitude at every $A$-site carbon atom. 
This flat band is made from a localized mode around 
the vacancy in each cell. 

We have obtained the band structures by the DFT-LDA model 
for $(3n\pm 1)\times(3n\pm 1)$ super cells with $n=2,3,4$ 
as well as $3n\times 3n$ cells with $n=2,3,4$. 
Always, we see the flat band, whose dispersion is essentially 
the same as the previous results. (Appendix~\ref{Appendix_DFT_LDA}) 
Interestingly, around the Fermi energy, 
the dispersion relation in the DFT-LDA modeling 
is always well reproduced by the $s$-$p$ tight-binding model. 
On an above mentioned finite size cell, 
we know that the $s$-$p$ model has a mapping to the $t$ model. 
Therefore, the flat-band mode coming from the pure topological origin of 
the $\pi$ network is identified as the zero mode. 
This is a reason for the terminology, which has already been used in 
Section~\ref{sec:level1}, and is a key for the next discussion. 

In an actual calculation with a $3n\times 3n$ super cell, 
hybridization between the Dirac mode 
and the zero mode appears significantly 
at the $\Gamma$ point of FBZ. 
However, even in this supercell, 
the flat band keeps its identity, when we choose the $k$ point 
apart from the $\Gamma$ point. 
Furthermore, there is a gap at the Dirac point for 
$(3n\pm 1)\times(3n\pm 1)$ supercells, at least when $n<O(10)$. 
(Appendix~\ref{Appendix_DFT_LDA})
Then, it is allowable to identify the flat band as zero modes in a finite-sized cell. 

When we consider mapping among single-particle eigen states, 
we use effective single-particle descriptions. 
In this representation, the zero mode might become a resonance in general, 
especially in the thermodynamic limit. 
For a test, we performed a finite-size scaling analysis of the flat band 
both in DFT and the $s$-$p$ tight-binding model. 
(See the DFT results in Appendix~\ref{Appendix_DFT_LDA}.)
Actually, the flat band is shown to be merged in 
the lower branches of the Dirac cone in the large cell size limit, 
when we adopt a linear scaling of the energy as a function of $1/n$ 
within a selected series of $(3n +1)\times(3n +1)$. 
The other series should show a similar behavior, 
when they are properly analized. 
(See also Appendix~\ref{Appendix_3n_mapping}.)

To have the good definition of the zero mode, in a thermodynamic limit, 
we adopt the next strategy. 
The strong correlation effect is taken into account in each 
finite-sized system, before approaching the limit. 
Having a well-identified correlated eigen state 
in an effective many-body model, we can have the limit by certifying 
identity of the correlated ground state of HGV. 
Thus, we consider a many-body model to describe correlation effects 
created by the special non-bonding damping mode, {\it i.e.} the zero mode. 

Here, we should comment on the eigen wavefunctions, $\psi_l({\bf r})$, 
of a single-particle part of the Hamiltonian. 
Using the set of  $\{\psi_l({\bf r})\}$ as an expanding basis, 
we have a full description of the many-body Hamiltonian. 
There, any unitary transformation among the basis does not change 
the final answer. We may make a good choice of the basis set 
to have a well-converged description.  
Therefore, we consider a representation obtained in a finite super cell 
to construct the scaling and the limit. 

The electron occupation of the zero mode 
in the many-body ground state is the second key. 
Let us consider neutral systems. 
For a finite super cell considered, the flat band is half-filled in DFT or TBMs. 
Furthermore, if we consider 
a cell with only one $V_{111}$ structure in the periodic boundary condition, 
one zero mode appears to be just half-filled.  
Namely, one electron occupies a single zero mode at the Fermi level 
in the non-interacting particle picture. 

Once the zero mode goes into a continuum of the Dirac modes, 
and if we consider the electron occupation in a free-particle picture, 
the zero mode has to be off from the half-filling. 
We should have a finite amplitude of 
the double occupation of the zero mode effectively in the electronic state, 
which implies that the correlation between two electrons occupying the zero mode 
becomes non negligible. 
Therefore, we have to consider the correlation effect carefully 
in the thermodynamic limit. 
To keep away from the energy increment by a finite amount of 
the double occupation amplitude of the localized mode, 
we need to take the large cell size limit, 
keeping identity of the zero mode 
from the continuum of the Dirac-cone modes. 

%%% Wannier transformation for the Hilbert subspace

To have a description of the local electron correlation in a DFT model, 
the Wannier transformation with the local projection operators 
is often applied.\cite{PhysRevB.56.12847,PhysRevB.65.035109,Wannier90} 
This method derives a mapping from the branches around the Fermi level of 
the DFT-LDA model to a tight-binding model. When the number of low-lying modes 
are selected to be the same, relevant eigen modes 
around the Fermi level are mapped to the corresponding modes of 
a simplified model, except for high-energy band crossing points. 

In the Wannier-transformation scheme, 
to have the complete expanding basis of the $\pi$ system, 
every local $\pi$ orbital, $\phi_i({\bf r})$, on the $i$-th carbon atom has to appear 
in the transformed representation. The size and the shape of $\phi_i({\bf r})$ 
is almost the same, irrespective of the location of the carbon site, $i$. 
Owing to this behavior, at least, 
the on-site interaction strength on $\phi_i({\bf r})$ is concluded to be 
approximately the same for all the carbon sites. 

The two-body effective interaction terms, {\it e.g.} 
the Hubbard interaction, are given as an operator expansion of 
the Coulomb scattering processes minus the mean-field counter 
terms.\cite{Hubbard1,Hubbard2,Hubbard3,PhysRevB.44.943} 
The interacting effective Hamiltonian along the line of the Wannierization 
takes the form of the extended Hubbard model with $N_C$ sites. 
The scattering processes in this representation 
should contain various off-site interactions, as well as the on-site 
Hubbard interaction $U$. 

Here, if one suppose that the longer-ranged interaction is irrelevant 
to be neglected, and if on-site repulsion $U$ only 
is kept in a general tight-binding description, 
the obtained Hubbard model might be less accurate for our target system. 
The description could be shifted from the true answer of 
the electronic structure in $V_{111}$. 
This is because we have a localized zero mode, 
in which the development of 
the electron correlation evolves differently from 
the extending quasi-particle modes in the Dirac cones. 
Note that the both modes, the zero mode 
and the quasi-particle modes, 
are given as combinations of $\phi_i({\bf r})$. 
But, the former is a localized state with exponentially decaying tails. 

Rather than the usage of $U$ on $\phi_i({\bf r})$, 
we should classify the scattering processes into 
an on-site interaction $U_0$ on the zero-mode and the others. 
Both of the scattering processes include 
the off-site interaction in representation 
of the LCAO picture using $\phi_i({\bf r})$. 
This statement is easily shown by expanding 
the $U_0$ term, which will be given in Eq.~(\ref{two_body_U}), into two-body 
effective interactions among several $\pi$ orbitals, $\phi_i({\bf r})$. 
Long-range interaction processes cannot be neglected. 

%%% Definition of effective many-body Hamiltonian 

Therefore, we go to the other way, starting from the band structure, 
or the momentum-space representation without 
using the Wannier transformation explicitly. 
Selecting only $(3n+1)\times(3n+1)$ super cells, 
we evaluate scattering amplitudes relevant for $V_{111}$. 
Here, we utilize the description by the multi-reference density functional 
theory (MR-DFT), which allows us to introduce two-body interaction terms 
as a quantum fluctuation 

in the energy density functional.\cite{Kusakabe,Kusakabe2} 
From now on, the creation and annihilation operators, 
$c^\dagger_{k,\sigma}$ and $c_{k,\sigma}$,  are to be defined 
by the basis given by the Kohn-Sham effective single-particle Hamiltonian. 
The charge density, $n({\bf r})$, is initially given by a local-density approximation 
(LDA), which is reconsidered after obtaining a solution of 
an effective many-body problem. 

%%% Evaluation of finiteness of U, J_direct, J_super

The correlation effect is counted by introducing 
a two-body correlation function explicitly in the energy functional. 
In the present situation, an important term is the next. 
Let us introduce a mean electron number in the zero mode, 
$N_0=\langle \hat{n}_{0\uparrow}+\hat{n}_{0\downarrow}\rangle$. 
At the half-fillling of the zero mode, $N_0=1$. 
We need to subtract a contribution from the Hartree term 
after consideration of the two-body term explicitly. Thus we have 
a kind of two-body effective Hamiltonian, $H_0^{(2)}$. 
\begin{equation}
H_0^{(2)} = \frac{U_0}{2} \left\{
:\left(\hat{n}_{0\uparrow}+\hat{n}_{0\downarrow}\right)^2:
-N_0 \left(\hat{n}_{0\uparrow}+\hat{n}_{0\downarrow}\right)\right\}.
\end{equation}
Here $:\hat{A}:$ represents a normal ordering of a product, $\hat{A}$, 
of the Fermion operators and the number operator is given as usual, 
$\hat{n}_{0\sigma}=c^\dagger_{k=0,\sigma}c_{k=0,\sigma}$. 
Under a condition, $N_0=1$, we have a 
corresponding interaction term appearing in the symmetric Anderson model, 
\begin{equation}
H_0^{(2)}= U_0\left(\hat{n}_{0\uparrow}
-\frac{1}{2}\right) 
\left(\hat{n}_{0\downarrow}
-\frac{1}{2}\right)
-\frac{U_0}{4},
\label{two_body_U}
\end{equation}
providing that the total system is minimized when the local charge neutral 
condition, $N_0 = 1$, is kept. 

One might consider a counter term coming from the 
exchange-correlation potential. In the discussion of $V_{111}$, 
we suppose that $N_0=1$ is kept by the correlation effect in a range of $\mu$. 
Actually, we can find a finite parameter space of the resulting Kondo Hamiltonian, 
where the $N_0=1$ condition is kept. (See Section~\ref{sec:level2_3-2}.)
The correlation causes redistribution of electrons in the Dirac cone. 
However, it appears only as a tiny change 
in a momentum distribution function along the energy axis. 
Within a reasonable accuracy, 
the final charge density is unchanged. This means that every functional of $n({\bf r})$ 
is not shifted after inclusion of the explicit correlation effects via the short-range 
interaction (correlation) terms in the effective Hamiltonian. 
Therefore, the contribution of the DFT exchange-correlation energy 
is not relevant for the present argument. 

\begin{figure}[htbp] 
\begin{center} 
\includegraphics[width=7cm]{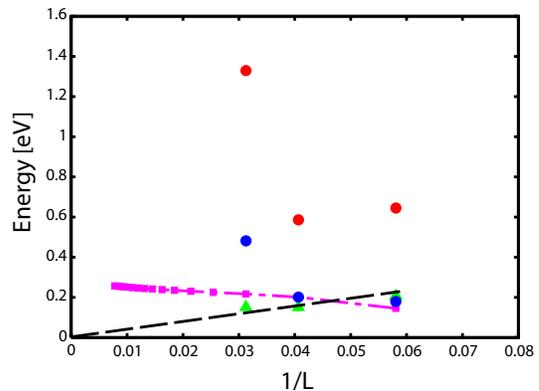} 
\end{center} 
\caption{Estimated interaction strength, $U_0$ (red dots), 
$J^{direct}$ (green triangles), and $J^{super}$ (blue dots). 
We adopt a $(3n+1)\times(3n+1)$ cell with $n=2,3,4$. The length 
$L$ of the super cell size is given by $L= (3n+1) \sqrt{3} a$ [\AA]
with $a$ [\AA] be the bond length of pure graphene. 
For $J^{direct}$, the dashed line gives a guide to the eyes. 
The magenta squares are $J^{super}$ estimated by the $t$ model, 
whose magnitudes are multiplied by a constant, 
so that the data for $n=3$ is adjusted to the DFT estimation. } 
\label{fig:Fig_Interaction_scale} 
\end{figure}

The value of correlation energy $U_0$ in Eq.~(\ref{two_body_U}) 
is estimated by evaluating it as a partially screened Coulomb integral. 
We utilized the DFT-LDA simulation to have the basis for the estimation. 
The plane-wave expansion with the norm-conserving pseudo-potentials 
is adopted with the cut-off of a value from 25 to 40~Ry. 
After checking reasonable convergence in the estimated scattering amplitudes, 
we utilize the value of 25~Ry for relatively large simulation cells. 
The estimation results in $U_0=0.6$, 0.6, 1.3 eV for 
$7\times7$, $10\times10$, and $13\times13$ cells, respectively. 
(Fig.~\ref{fig:Fig_Interaction_scale})
Although the value is not converging systematically, we suppose that 
there is a finite value of $U_0$ of order eV. 

In addition to Eq.~(\ref{two_body_U}), 
we have many types of scattering processes 
promoting magnetic behavior. 
Owing to finiteness of $U_0$, we assume $N_0=1$ at 
around the charge neutral point, {\it i.e.} $\mu=0$. 
Among processes, we keep the spin-dependent scatterings 
in a form of the $s$-$d$ interaction in our discussion. 
Similar estimation of the spin-exchange scattering with 
the partial screening tells us the next result. 

Let us now consider amplitude $J_{kk'}$ 
of the spin-dependent scattering. Here the sufix $k$ 
may be interpreted as a combined index of the band and the wave vector. 
In our formalism, however, we utilize a basis of modes defined in a 
selected large super cell. Therefore, $k$ specifies a low-energy $\pi$ mode. 
Note that the matrix element, $J_{kk'}$, represents a scattering amplitude 
of an electron in a mode $k$ to another mode $k'$ by 
a magnetic scattering by the zero mode spin. 

As the lowest order direct exchange scattering, $J^{direct}_{kk'}$, 
we have always the diagonal ferromagnetic contribution, $J^{direct}_{kk}$. 
This is only because the zero mode and each Dirac mode 
are orthogonal with each other. The off-diagonal terms in the first order 
are relatively smaller than the diagonal element. 
Because the Dirac mode is extending wave, while the zero mode is 
localized, the absolute value of $J^{direct}_{kk'}$ decreases, 
when the super cell size increases. 
We have actually a scaling relation of $J^{direct}_{kk'} \sim O(1/n)$, 
which goes to zero in the large cell limit. 
Therefore, the exchange integral $J^{direct}_{kk'}$ should behave 
as a monotonically decreasing function of the cell size. 
(Fig.~\ref{fig:Fig_Interaction_scale}) 

There is a higher-order magnetic scattering term. 
A typical second order contribution reads, 
\begin{eqnarray}
\label{J_super}
J^{super}_{kk'}
&=&\frac{1}{2}\sum_{|\epsilon_p|>\epsilon_c}\left\{
\frac{V_{(0k)\rightarrow(0p)}V_{(0p)\rightarrow(k'0)}}
{\varepsilon_p-\varepsilon_k} \right. \nonumber \\
&&+\left. \frac{V_{(0k)\rightarrow(p0)}V_{(p0)\rightarrow(k'0)}}
{\varepsilon_p-\varepsilon_k} \right\}.
\end{eqnarray}
Here, $\varepsilon_k$ denotes the single particle energy of the mode $k$, 
where $\varepsilon_k$ and $\varepsilon_{k'}$ are within a cut off of $\varepsilon_c$, 
and $V_{(0k)\rightarrow(pq)}$ is a Coulomb scattering amplitude. 
For details of derivation, see Appendix~\ref{Appendix_4n_superexchange}. 

Let's suppose that we have a Born-von-Karman boundary condition 
for the description of the wave function of the Dirac bands of graphene. 
Wave functions in the Dirac cones are characterized by 
a wave vector around the K or K' points. 
When the difference of ${\bf k}$ from ${\bf k}_{K}$ (or ${\bf k}_{K'}$) 
is small, we have an asymptotic wave function, 
{\it i.e.} those of the $k \cdot p$ approximation in the Weyl representation. 
Introducing the envelop function, we have the long-wave length limit 
for the low energy excitations. 
In this limit, the exchange scattering within each valley 
should have a momentum independent amplitude. 

For $V_{111}$, we have unbalanced sublattice structure. Therefore, 
the so-called chiral symmetry in the pure graphene is not used. 
Furthermore, we evaluate the amplitudes of two-particle scattering among 
modes at the $\Gamma$-point in FBZ. Therefore, $k$ represents the band index. 
Namely, without using the Born-von-Karman boundary condition, 
we perform a size-scaling analysis. 
The value and the sign of $J^{super}_{kk'}$ might 
depend on the cell size, again. 

In the large cell limit, we should have amplitudes 
which are almost $k$-independent in the low energy regime except for 
a degrees of freedom coming from the sublattice site index. 
In this limit, the form of Eq.~(\ref{J_super}) can have a non-negligible 
contribution owing to the integration with respect to the intermediate modes. 
In addition, similar to the old super-exchange mechanism, 
anti-ferromagnetic contribution should be dominant, because of 
a larger domain for the intermediate state 
than that for ferromagnetic processes. 

Material dependence of $J^{super}_{kk'}$ comes from 
the sign and the magnitude 
of amplitude, $V_{(0k)\rightarrow(0p)}$ and $V_{(0p)\rightarrow(k'0)}$, 
and the spectrum $\varepsilon_k$. 
In the present case, we have relatively large contribution 
for the anti-ferromagnetic channel, where 
$V_{(0k)\rightarrow(0p)}V_{(0p)\rightarrow(k'0)}$ is positive real valued. 

In our estimation, we see a systematic increase of relative contribution 
in the anti-ferromagnetic channels for larger cells. 
The order of magnitude for $J^{super}_{kk'}$ is estimated to be 
$\sim 0.2$ eV in finite cells, 
whose temperature scale is $O(10^3)$ K. 
(Fig.~\ref{fig:Fig_Interaction_scale})
However, the convergence is not enough achieved 
within an accessible cell size for the DFT simulations. 

To show converging behavior of the asymptotic form of 
$J^{super}_{kk'}$, we combine results by the $t$ model 
with the DFT estimation. (See Appendix~\ref{Appendix_4n_superexchange}.) 
In Fig.~\ref{fig:Fig_Interaction_scale}, 
the total magnitude of $J^{super}$ is multiplied by a factor 
to adjust the two values for a cell of $n=3$. 
In comparison with $J^{super}$, the direct exchange 
$J^{direct}$ is negligible in the large cell size limit. 

\subsection{\label{sec:level2_2-1}Representation in a Kondo model}

The effective model of $V_{111}$ 
is represented by a Kondo Hamiltonian below.
Here, $\epsilon_0 = \varepsilon_0 -U_0/2 -\mu$, 
and $\epsilon_k = \varepsilon_k-\mu $. \\

\begin{eqnarray}
\label{eq:H_Kondo}
H_{Kondo}&=&H_0+H_J+H_{D},\\
%\label{eq:H_0}
H_0
&=&\sum_{\sigma = \uparrow \downarrow} 
\epsilon_0 c^{\dag}_{0\sigma}c_{0\sigma}
+U_0 \hat{n}_{0\uparrow}\hat{n}_{0\downarrow},
\nonumber\\
%\label{eq:H_J}
H_J
&=&(c^{\dag}_{0\sigma} \vec{\sigma}_{\sigma \sigma '} c_{0\sigma '})
\sum_{kk'}J_{kk'}(c^{\dag}_{k\tau} \vec{\sigma}_{\tau \tau '} c_{k'\tau '}),
\nonumber\\
%\label{eq:H_Dirac}
H_{D}&=&\sum_{k, \tau} \epsilon_k c^{\dag}_{k\tau}c_{k\tau}.\nonumber
\end{eqnarray}
In Eqs. (\ref{eq:H_Kondo}), $H_0$ represents the energy level of the zero mode 
and the on-site Coulomb repulsion $U_0$ on the zero mode. 
$H_J$ represents the effective scattering process of conduction electrons 
caused by exchange interactions $J_{kk'}$ 
between the electrons on the zero mode and the Dirac electrons. 
$H_{D}$ is the Hamiltonian for the Dirac electrons. 
$\sigma$ and $\tau$ indicate the spin components of electrons 
on the zero mode and the Dirac electrons, respectively. 
Repeated indexes, $\sigma$ and $\tau$, in $H_J$ are 
assumed to be summed with respect to the spin direction. 

As discussed in Section~\ref{sec:level2_2-0}, 
the simplest modeling of $V_{111}$ incorporating the correlation effect 
is to have the diagonal single-body term and 
the off-diagonal general two-body terms. 
Among the two-body scatterings, the on-site interaction 
at the zero mode and the exchange scattering processes 
keeping the electron number of the zero mode are relevant 
for the strong correlation regime. 
Therefore, the above modeling is 
a natural model for HGV with topological zero modes. 

%%%
\subsection{\label{sec:level2_2-2}Transformation to an impurity Anderson model}

In order to treat the Kondo hamiltonian numerically, 
we transform $H_{Kondo}$ into an effective Anderson model.\cite{PhysRev.124.41} 
We can convert the form of $H_J$ in Eqs.~(\ref{eq:H_Kondo}) 
into a separable form by diagonalizing $J_{kk'}$ with a unitary matrix as

\begin{eqnarray}
\label{eq:H_J_sep}H_J&=&(c^{\dag}_{0\sigma} \vec{\sigma}_{\sigma \sigma '} c_{0\sigma '})\sum_{kk'}J_{kk'}(c^{\dag}_{k\tau} \vec{\sigma}_{\tau \tau '} c_{k'\tau '})\nonumber\\
&=&(c^{\dag}_{0\sigma} \vec{\sigma}_{\sigma \sigma '} c_{0\sigma '}) \nonumber \\
&\times&
\sum_{l}\Bigl\{{(\sum_k  c^{\dag}_{k\tau} U_{J,kl})}
\bar{J}_l\vec{\sigma}_{\tau \tau '}
{(\sum_{k'} U_{J,lk'}^{-1} c_{k'\tau '})}\Bigr\}\nonumber \\
&=&(c^{\dag}_{0\sigma} \vec{\sigma}_{\sigma \sigma '} c_{0\sigma '})
\sum_{l} \bar{c}^{\dag}_{l\tau}
\bar{J}_l\vec{\sigma}_{\tau \tau '} \bar{c}_{l\tau'}.
\end{eqnarray}
Here, $\bar{J}_l$ are matrix elements of the diagonalized matrix, 
$\bar{J}=U_J^{-1} J U_J$, and $\bar{c}^{\dag}_{l\tau}$ 
(or $\bar{c}_{l\tau '}$) are given by $\sum_k c^{\dag}_{k\tau} U_{J,kl}$ 
(or $\sum_{k'} U_{J,lk'}^{-1} c_{k'\tau '}$). 
We assume that $\bar{J}_l$ is ordered in a descending series. 

By considering the most effective $\bar{J}_{l=1}$ alone, 
and by 
obtaining $\bar{H}_{D}$ by substituting $c_{k\tau}$
in an expansion with $\bar{c}_{k\tau}$ for $H_{D}$ 
in Eqs.~(\ref{eq:H_Kondo}), 
we can transform $H_{Kondo}$ into $H_{Anderson}$ as 
\begin{eqnarray}
%\label{eq:H_K_A}
H_{Kondo}&=&H_{Anderson}+O(1/N_C),\nonumber\\
\label{eq:H_Anderson}
H_{Anderson}&=&\bar{H}_0 + \bar{H}_{D}+\bar{H}_{hyb}. 
\end{eqnarray}
Here, each term is given by, 
\begin{eqnarray}
%\label{eq:tilde_H_0}
\bar{H}_0&=&\sum_{\sigma=\uparrow\downarrow}\epsilon_0c^{\dag}_{0\sigma}c_{0\sigma}
+U_0\hat{n}_{0\uparrow}\hat{n}_{0\downarrow}\nonumber\\
&+&\sum_{\sigma=\uparrow\downarrow}{\epsilon}_{11}'\bar{c}^{\dag}_{1\sigma}{\bar{c}_{1\sigma}}+\bar{J}_{1}(c^{\dag}_{0\sigma}\vec{\sigma}_{\sigma\sigma'}c_{0\sigma'})(\bar{c}^{\dag}_{1\tau}\vec{\sigma}_{\tau\tau'}\bar{c}_{1\tau'}),\nonumber
\end{eqnarray}
\begin{eqnarray}
%\label{eq:tilde_H_Dirac}
\bar{H}_{D}&=&\sum_{k\neq 1,\sigma}{\epsilon}_{kk}' \bar{c}^{\dag}_{k\sigma}\bar{c}_{k\sigma},\nonumber\\
%\label{eq:H_hyb}
\bar{H}_{hyb}&=&\sum_{k\neq 1, \sigma}
(\epsilon_{k1}'\bar{c}^{\dag}_{k\sigma}\bar{c}_{1\sigma}+H.c.). \nonumber
\end{eqnarray}
Note that hybridization $\epsilon_{k1}'$ 
as single-body parts in Eqs.~(\ref{eq:H_Anderson}) 
come from the off-diagonal elements of 
$\epsilon_{kk'}' =\sum_{k''} U_{J,kk''}^{-1} \epsilon_{k''} U_{J,k''k'}$.
%%%%%%%%%%%%%%%%%%%%%%%%%%%%%%%%%%%%%%%%%%%%%%%%%%%%%%%

\subsection{\label{sec:level2_2-3}The $s$-channel exchange term}

In the transformation in Section~\ref{sec:level2_2-2}, we note that 
there could be several $\bar{J}_{l}$s, 
which became non-negligible, in general. 
Besides, $U_J$ does give various off-diagonal hybridization terms 
in addition to the diagonal term, $\bar{H}_{D}$. 
Let's evaluate these terms by considering an ideal case. 

As discussed in Section~\ref{sec:level2_2-0}, $J_{k,k'}$ approaches to a 
constant asymptotically for the low-energy quasi-particle modes, 
when the system size becomes enough large. 
Therefore, the most typical model of $J_{k,k'}$ is 
a matrix with constant elements. 

Introducing an $n\times n$ matrix, $J=(J_{k,k'})$, as 
\begin{equation}
  J = J_0 \left(
    \begin{array}{cccc}
      1 & 1 & \cdots&1 \\
      1& 1 & \cdots&1 \\
      \vdots & \vdots & \ddots&\vdots\\
      1&1&\cdots&1
    \end{array}
  \right),
\label{Model_J}
\end{equation}
we see that $\bar{J}_{l=1}=n J_0$ and $\bar{J}_{l>1}=0$. 
To diagonalize $J$, we can choose a unitary matrix 
 \[
  U_J = \left(
    \begin{array}{ccccc} \displaystyle
      1/\sqrt{n} & 1/\sqrt{2} &1/\sqrt{6}& \cdots&1/\sqrt{n(n-1)} \\
      1/\sqrt{n}& -1/\sqrt{2} &1/\sqrt{6}& \cdots&1/\sqrt{n(n-1)} \\
      1/\sqrt{n}& 0 &-2/\sqrt{6}& \cdots&1/\sqrt{n(n-1)} \\
      \vdots & \vdots & \vdots&\ddots&\vdots\\
      1/\sqrt{n}&0&0&\cdots&\frac{-(n-1)}{\sqrt{n(n-1)}}
    \end{array}
  \right).
\]
Indeed, the transformed $J$ is, 
  \[
  U_J^{-1}JU_J = J_0 \left(
    \begin{array}{cccc}
      n & 0 & \cdots&0 \\
      0& 0 & \cdots&0 \\
      \vdots & \vdots & \ddots&\vdots\\
      0&0&\cdots&0
    \end{array}
  \right).
\]
By applying $U_J$, the single-particle part is transformed. 
Actually, the unitary transformation changes the diagonal matrix, 
\[
  \epsilon = \left(
    \begin{array}{cccc}
      \epsilon _1 & 0 & \cdots&0 \\
      0& \epsilon _2 & \cdots&0 \\
      \vdots & \vdots & \ddots&\vdots\\
      0&0&\cdots&\epsilon _n
    \end{array}
  \right)
\]
into
  \[
  \epsilon '=U_J^{-1}\epsilon U_J = \left(
    \begin{array}{cccc}
      \epsilon '_{11} & \epsilon '_{12} & \cdots&\epsilon '_{1n} \\
      \epsilon '_{21}& \epsilon '_{22} & \cdots&\epsilon '_{2n} \\
      \vdots & \vdots & \ddots&\vdots\\
      \epsilon '_{n1}&\epsilon '_{n2}&\cdots&\epsilon '_{nn}
    \end{array}
  \right).
\]
Let us introduce another transformation to have an ideal form. 
Taking a $(n-1)\times(n-1)$ unitary matrix $u_{\epsilon'}$ 
which diagonalizes a block matrix
  \[
  \epsilon ''= \left(
    \begin{array}{cccc}
      \epsilon '_{22} & \epsilon '_{23} & \cdots&\epsilon '_{2n} \\
      \epsilon '_{32}& \epsilon '_{33} & \cdots&\epsilon '_{3n} \\
      \vdots & \vdots & \ddots&\vdots\\
      \epsilon '_{n2}&\epsilon '_{n3}&\cdots&\epsilon '_{nn}
    \end{array}
  \right),
\]
we can make a unitary matrix $U_{J\epsilon}=U_JU_\epsilon$ where
  \[
  U_\epsilon= \left(
    \begin{array}{cc}
      1 & 0 \\
      0& u_{\epsilon '} 
    \end{array}
  \right).
\]
The transformation by $U_{J\epsilon}$ keeps the diagonal form of the $J$ term. 
  \begin{eqnarray}
\tilde{J}&=&
  U_{J\epsilon}^{-1}JU_{J\epsilon} \nonumber \\ 
&=& J_0 \left(
    \begin{array}{cccc}
      n & 0 & \cdots&0 \\
      0& 0 & \cdots&0 \\
      \vdots & \vdots & \ddots&\vdots\\
      0&0&\cdots&0
    \end{array}
  \right)
=\left(
    \begin{array}{cccc}
      \tilde{J}_{l=1} & 0 & \cdots&0 \\
      0& 0 & \cdots&0 \\
      \vdots & \vdots & \ddots&\vdots\\
      0&0&\cdots&0
    \end{array}
  \right),
\end{eqnarray}
while the $\epsilon$ term changes its shape into a desired form as, 
  \[
  \tilde{\epsilon}=U_{J\epsilon}^{-1}\epsilon U_{J\epsilon} = \left(
    \begin{array}{ccccc}
     \tilde{\epsilon}_{1} &V_{21}^*&V_{31}^*& \cdots&V_{n1}^*\\
      V_{21}& \tilde{\epsilon}_{2}&0& \cdots&0\\
      V_{31}& 0 &\tilde{\epsilon}_{3}& \cdots&0 \\
      \vdots & \vdots & \vdots&\ddots&\vdots\\
      V_{n1}&0&0&\cdots&\tilde{\epsilon}_{n}
    \end{array}
  \right). 
\]
We see that the spectrum of $\epsilon_i$ is almost reproduced by 
$\tilde{\epsilon}_{i}$, when $n$ is enough large. 
Therefore, the Anderson Hamiltonian shown in Section~\ref{sec:level2_2-2} 
represents the Kondo Hamiltonian without any essential correction, 
if $J_{k,k'}$ behaves as a constant among relevant quasi-particle modes. 

In case when $\epsilon_i$ has degeneracy, we have additional 
simplification in 
the transformed matrices, $\tilde{\epsilon}$ and $\tilde{J}$. 
This comes from the $s$-wave nature of the present model 
interaction, $H_J$, with $J$ by Eq.~(\ref{Model_J}). 
To see this behavior, let assume that $\epsilon_i=\epsilon_m$ 
for $1\le i\le m$. Let us consider a transformation among 
degenerate states by, 

 \begin{eqnarray}
  U_P &=& \left( \begin{array}{cc} U_{J,m} & 0 \\ 0 & I \end{array} \right),\nonumber \\
  U_{J,m} &=& \left(
    \begin{array}{cccc}
      1/\sqrt{m} & 1/\sqrt{2} & \cdots&1/\sqrt{m(m-1)} \\
      1/\sqrt{m}& -1/\sqrt{2} & \cdots&1/\sqrt{m(m-1)} \\
      1/\sqrt{m}& 0 & \cdots&1/\sqrt{m(m-1)} \\
      \vdots & \vdots &\ddots&\vdots\\
      1/\sqrt{m}&0&\cdots&-(m-1)/\sqrt{m(m-1)}
    \end{array}
  \right).\nonumber
\end{eqnarray}
Here $I$ is an $(n-m)\times(n-m)$ identity matrix. 
Introducing another $(n-m)\times(n-m)$ matrix, 
\[  J_{n-m} = J_0 \left(
    \begin{array}{cccc}
      1 & 1 & \cdots&1 \\
      1 & 1 & \cdots&1 \\
      \vdots & \vdots & \ddots&\vdots\\
      1 & 1 & \cdots&1
    \end{array}
  \right),
\]
we see that, 
\[U_P^{-1} J U_P 
= \left( \begin{array}{cc} U_{J,m}^{-1}J_{m}U_{J,m} & 
\hat{J}_{m} \\ ^t\hat{J}_{m} & J_{n-m} \end{array} \right),\]
\[U_{J,m}^{-1}J_{m}U_{J,m}=J_0 \left(
    \begin{array}{cccc}
      m & 0 & \cdots&0 \\
      0& 0 & \cdots&0 \\
      \vdots & \vdots & \ddots&\vdots\\
      0&0&\cdots&0
    \end{array}
  \right),
\]
\[\hat{J}_{m}=J_0 \left(
    \begin{array}{cccc}
      \sqrt{m} & \sqrt{m} & \cdots&\sqrt{m} \\
      0& 0 & \cdots&0 \\
      \vdots & \vdots & \ddots&\vdots\\
      0&0&\cdots&0
    \end{array}
  \right),
\]
\[U_P^{-1} \epsilon U_P = \epsilon,\]
The above result is naturally interpreted using 
representations of the permutation group of $m$ elements. 
Namely, only the fully symmetric representation has the 
exchange contribution, while the other representations have zero 
matrix elements, as far as $J$ is assumed to be given by Eq.~(\ref{Model_J}). 
Keeping the symmetric representation, 
omitting the others in the degenerate levels, we have a reduced 
exchange matrix, 
\[
  J^{red} = J_0 \left(
    \begin{array}{cccc}
      m & \sqrt{m} & \cdots&\sqrt{m} \\
      \sqrt{m} & 1 & \cdots&1 \\
      \vdots & \vdots & \ddots&\vdots\\
      \sqrt{m} &1&\cdots&1
    \end{array}
  \right),
\]
where $J^{red}$ has no more constant matrix elements. 
Therefore, we can choose each symmetric representation 
created by degenerate states, which may be called states of 
the $s$ symmetry, to construct the reduced exchange matrix. 

When the matrix $J$ is given as a matrix with constant elements, 
our Hamiltonian reads as, 
\begin{eqnarray}
H_{Kondo}&=&H_{Anderson},\nonumber\\
H_{Anderson}&=&\tilde{H}_0 + \tilde{H}_{D}+\tilde{H}_{hyb}. 
\label{TIAM}
\end{eqnarray}
Here, each term is given by, 
\begin{eqnarray}
\tilde{H}_0&=&\sum_{\sigma=\uparrow\downarrow}\epsilon_0c^{\dag}_{0\sigma}c_{0\sigma}
+U_0\hat{n}_{0\uparrow}\hat{n}_{0\downarrow}\nonumber\\
&+&\sum_{\sigma=\uparrow\downarrow}\tilde{\epsilon}_{1}\tilde{c}^{\dag}_{1\sigma}{\tilde{c}_{1\sigma}}+\tilde{J}_{1}(c^{\dag}_{0\sigma}\vec{\sigma}_{\sigma\sigma'}c_{0\sigma'})(\tilde{c}^{\dag}_{1\tau}\vec{\sigma}_{\tau\tau'}\tilde{c}_{1\tau'}),\nonumber \\
\tilde{H}_{D}&=&\sum_{k\neq 1,\sigma}\tilde{\epsilon}_{k} \tilde{c}^{\dag}_{k\sigma}\tilde{c}_{k\sigma},\nonumber\\
\tilde{H}_{hyb}&=&\sum_{k\neq 1, \sigma}
(V_{k1}\tilde{c}^{\dag}_{k\sigma}\tilde{c}_{1\sigma}+H.c.). \nonumber
\end{eqnarray}
Here, $\tilde{c}^{\dag}_{l\tau}$ 
(or $\tilde{c}_{l\tau '}$) are given by 
$\sum_k c^{\dag}_{k\tau}U_{J\epsilon,kl}$ 
(or $\sum_{k'} U_{J\epsilon,lk'}^{-1} c_{k'\tau '}$). 
The one-body Hamiltonian, $\tilde{H}_{D}$, describes 
renormalized Dirac modes with the modified spectrum, $\tilde{\epsilon}_k$. 

%%%%%%%%%%%%%%%%%%%%%%%%%%%%%%%%%%%%%%%%%%%%%%%%%%%%%%%
\section{\label{sec:level1_3}Calculation results and discussions}

\subsection{\label{sec:level2_3-1}The CT-HYB-QMC method}
In this section, we apply a numerical method for an impurity-problem to 
the Anderson Hamiltonian derived in the last section. 
We can trace the behavior of $H_{Anderson}$ with a continuous-time 
hybridization-expansion matrix solver 
(the CT-HYB-QMC solver) of the quantum Monte-Carlo method. 
The solver is provided by the TRIQS project.\cite{TRIQS} 

A reason for the application is its powerful ability  
to handle the impurity problem with a few impurity sites. In addition, 
the QMC algorithm is useful to treat a general bath Green function. 
The temperature range in our concern is accessible by CT-HYB-QMC. 
Another reason is because the CT-HYB algorithm treats any 
interacting impurity model 
with finite interaction strength of $U_0$ and $J_{k,k'}$ 
effectively.\cite{PhysRevLett.97.076405,PhysRevB.74.155107} 
We applied a transformation to the Anderson model, which allows us to apply 
the package with CT-HYB-QMC directly. 

We regard the effective Anderson model described in Eq.~(\ref{eq:H_Anderson}) 
or Eq.~(\ref{TIAM}) 
as a two-site impurity Anderson model (TIAM). 
The center orbital is the zero mode, $\phi_0({\bf r})$. 
As a result of the transformation in Section~\ref{sec:level2_2-2}, 
we have an additional localized orbital. 
Let us name an orbital, 
which is magnetically hybridized with the zero mode 
by the exchange interaction of $\tilde{J}_{l=1}$, 
`the first orbital'. As usual impurity Anderson systems, 
we have hybridization terms, $\tilde{H}_{hyb}$, which form `a Kondo cloud' 
in the renormalized Dirac modes. 
We calculated the mean value of the number of electrons in `the first orbital', $N_{first}=\average{\hat{n}_{1\uparrow}}+\average{\hat{n}_{1\downarrow}}$ where $\hat{n}_{1\sigma}=\tilde{c}^{\dag}_{1\sigma}{\tilde{c}_{1\sigma}}$, 
as well as that in the zero mode, $N_{zero}=\average{\hat{n}_{0\uparrow}}+\average{\hat{n}_{0\downarrow}}$,  
to characterize the electronic state and the correlation effect in the system. 

\subsection{\label{sec:level2_3-2}Calculation results}

As a consequence of the unitary transformation on the Dirac modes, 
an energy level $\tilde{\epsilon}_1$, which is the site energy of the first orbital, 
appears at the Dirac point. 
So, we take $\epsilon_0=-U_0/2-\mu$ for the zero mode 
and $\tilde{\epsilon}_1=-\mu$ for the first orbital in this calculation.   

The density of states in the low-energy excitations of Dirac electrons 
are modeled by that of the $t$ model. We adopted a value 
$t=3.03$[eV] for the nearest neighbor transfer.\cite{PhysRevB.46.1804}  
In this section, we take the inverse temperature 
$\beta = 100/t = 33$[1/eV]. 
The interaction strength of $U_0=1$ [eV] is adopted. 

We assume that each element $J_{k,k'}$ of the effective exchange 
scattering matrix $J=(J_{k,k'})$ is almost a constant among 
the low-energy modes in the Dirac cone, and its size $|J|=\tilde{J}_{l=1}$ is 
in a range estimated in Sec.~\ref{sec:level2_2-0}. 
Therefore, we treat $\tilde{J}_{l=1}$ as a parameter within this energy scale. 

Another parameter for this modeling 
is the energy range of Dirac modes, for which $J_{k,k'}$ is finite. 
This value corresponds to the matrix size $n$ of $J$. 
We have checked that the result is qualitatively the same 
irrespective of the range of Dirac modes for which $V_{k1}$ is finite. 
Thus, we adopt a continuous representation of $\tilde{\epsilon}_k$ and $V_{k1}$, 
where an average, $V_0$, of $V_{k1}$ is taken for simplicity. 
We will discuss dependence of the main results 
on $V_{k1}$, which is assumed to be in a scale of $J_0$ in simulations. 

\begin{figure}[htbp] 
\begin{center} 
\begin{flushleft}(a)\end{flushleft} \vspace*{0mm}
\includegraphics[width=9cm]{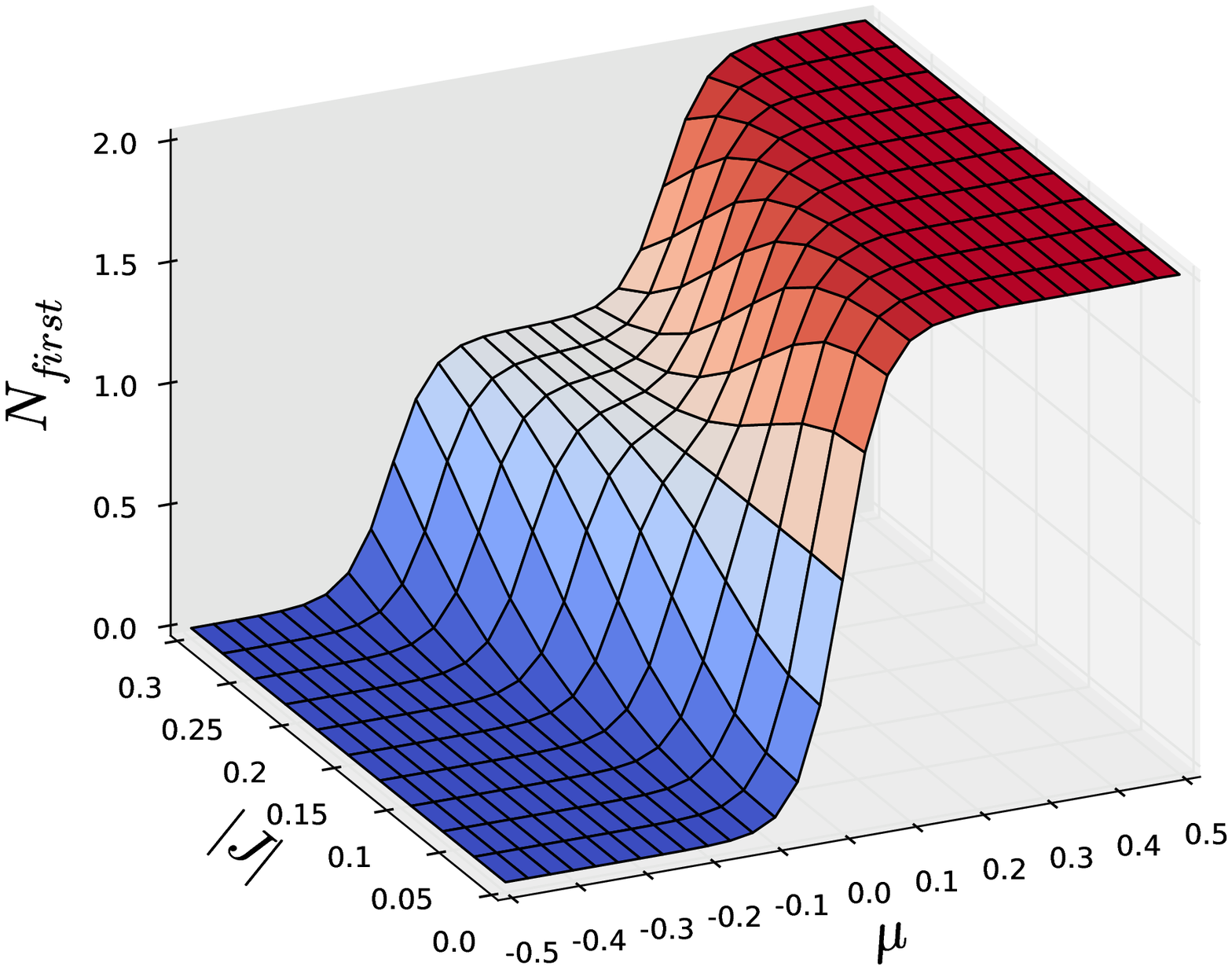} 
\end{center}
\begin{center}
\begin{flushleft}(b)\end{flushleft} \vspace*{0mm}
\includegraphics[width=9cm]{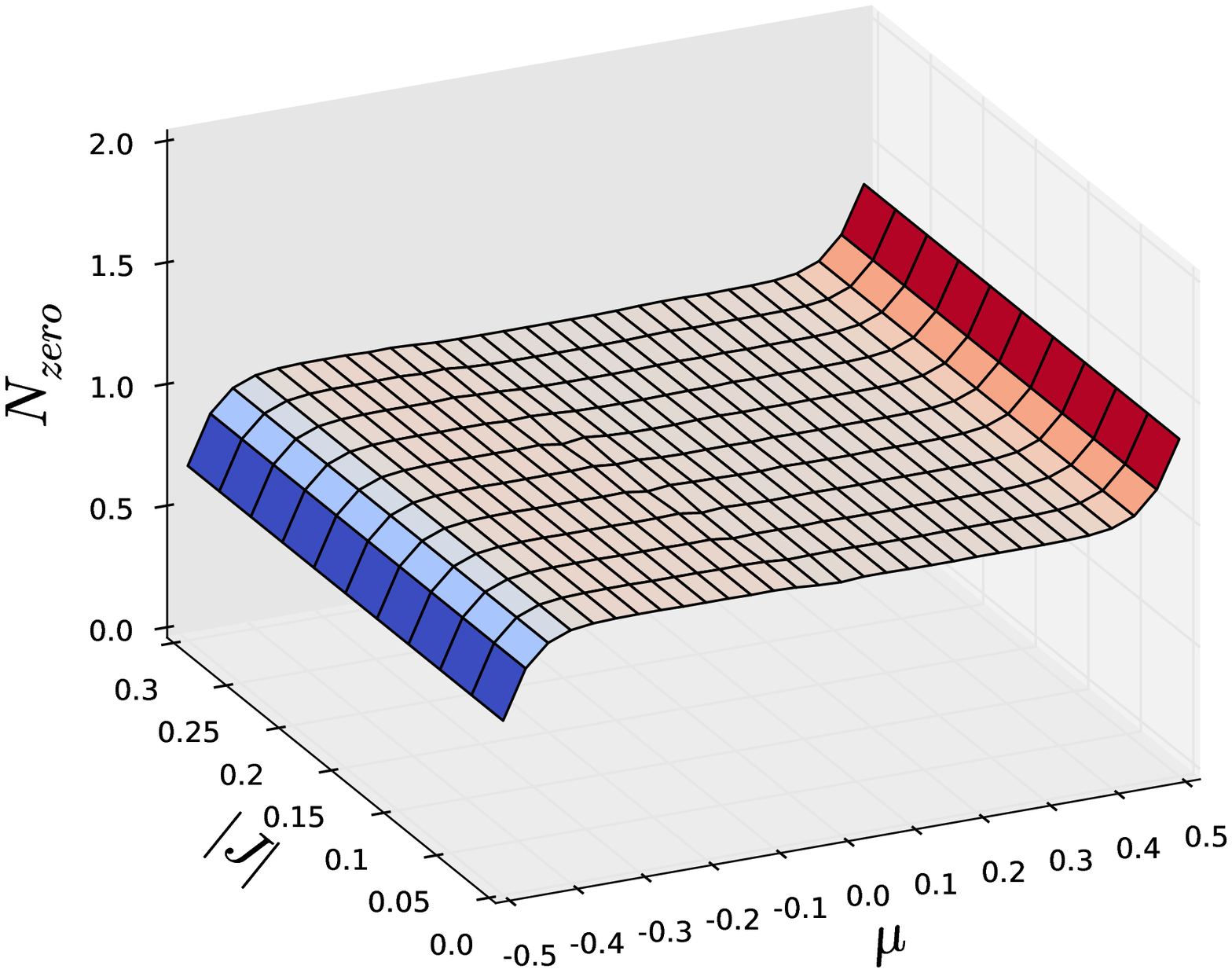} 
\end{center} 
\caption{(a) The number of electrons in 
``the first orbital'', $N_{first}$, and its 
chemical potential dependence. 
The chemical potential is given by $\mu$ [eV]. 
The exchange interaction, $|J|$, between the zero mode and 
the Dirac electrons is anti-ferromagnetic. 
We can see continuous change of $N_{first}$ where $|J|$ is small, 
and stepwise change of $N_{first}$ where $|J|$ is large. 
(b) The number of electrons in the zero mode, $N_{zero}$. 
When $|\mu|$ is smaller than $U_0/2=0.5$ [eV], 
$N_{zero}\simeq 1$, and the zero mode is kept at the half-filling.} 
\label{fig:Fig.1} 
\end{figure}

The chemical potential $\mu$ changes the Fermi level in the Dirac electrons, 
which affects the occupation of localized modes in turn. 
The chemical potential dependence of $N_{first}$ (and $N_{zero}$) 
of the model is obtained and shown in Fig.~\ref{fig:Fig.1}. 

In our model, the on-site interaction $U_0$ at the zero mode 
has enough large value of O(eV). 
Therefore, the zero mode keeps its identity as a singly occupied mode. 
Namely, $N_0=N_{zero}$ is almost kept at 1, at least if $|\mu| < U/2$, 
where the condition of $N_0=1$ is necessary 
for the consistency as a simulation of MR-DFT. 
Then, an unpaired spin exists at the zero mode. 
The problem described by $H_{Anderson}$ is magnetic 
screening effect mediated by exchange interactions. 

When $|J|$ is smaller than about 0.1[eV], $N_{first}$ 
drops from 0.0 to 2.0 instantly. 
The result tells that electron occupation at 
an effective localized level, {\it i.e.} the first orbital, 
follows the un-correlated free electron behavior. 
This is a weak-coupling limit of the system, 
although $U_0$ keeps the half-filling of the zero mode. 
Therefore, when $\tilde{J}_{l=1}$ is too small 
against other parameters, {\it i.e.} the temperature and $V_{k1}$, 
we have a weak-coupling scheme without magnetic screening, 

When $\tilde{J}_{l=1}$ is large enough, 
anti-ferromagnetic correlation evolves between 
the zero mode and the effective localized level, {\it i.e.} the first orbital. 
Indeed, $N_{first}$ shows a stepwise behavior. 
This behavior indicates that correlation develops in the Dirac electrons. 
This result suggests that $J_{kk'}$ in Eqs. (\ref{eq:H_Kondo}) can 
mediate the Kondo screening of the spin 1/2 in the zero mode 
at the low temperature. 
The region for this screening to work has a V-shaped structure 
opening in the larger $|J|$ region widely against $\mu$ in the $\mu$-$|J|$ space. 

Here $V_{k1}$ is assumed to have a value of order $0.1$ [eV]. 
When $V_{k1}$ is getting smaller, the graph shown in 
Fig.~\ref{fig:Fig.1} becomes much sharper at the edges of the V-shaped structure. 
Therefore, we conclude that the physical picture given here is qualitatively 
kept as far as $|J|$ being finite and $V_{k1}$ is small. 

\subsection{\label{sec:level2_3-3}Strong coupling regime in the large system size limit}

\begin{figure}[htbp]
\begin{center}
\includegraphics[width=8cm]{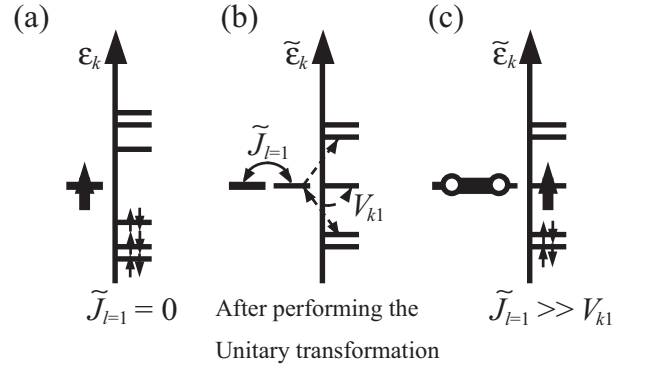}
\end{center}
\caption{Schematic viewgraph of crossover happened 
for scattering-driven Kondo screening. (a) A center level by the thick line 
represents a zero mode of $V_{111}$ and the other levels are from 
the Dirac cone. When $J_0$ is negligible, 
the zero mode with $U_0>0$ has a single free spin. 
(b) The two-body scattering 
processes by $J_{k,k'}$ mediate the relevant super exchange $J_0$, 
where a unitary transformation given by $J_{k,k'}$ for the metallic Dirac cone states 
selects the first orbital $\phi_1({\bf r})$, 
defines renormalized spectrum $\tilde{\varepsilon}_i$, 
and determines one-body hybridization $V_{1,k}$. 
(c) When $\tilde{J}_{l=1}$ is kept finite, the screening happens, 
while $V_{21}$ approaches zero, if the system size increases, 
providing the renormalized Dirac cone.}
\label{fig:Doublet}
\end{figure}

The physical picture of the above result may be interpreted as follows. 
Consider a finite anti-ferromagnetic coupling $\tilde{J}_{l=1}$. 
This interaction creates or defines an imaginative localized state 
as a linear combination of the Dirac levels. (Fig.~\ref{fig:Doublet})
This interaction-driven localized orbital, {\it i.e.} the first orbital, 
is also half-filled, when $\mu$ is at the Dirac point. 
An electron in this mode and the zero-mode electron couple 
with the anti-ferromagnetic interaction. 
Owing to a finite $V_{k1}$, however, the screening becomes 
imperfect. A finite amplitude of the magnetic moment 
may appear at the zero mode. 

In our model, thus, we have 
a weak coupling limit of $\tilde{J}_{l=1}=0$ (Fig.~\ref{fig:Doublet}~(a))
where an unscreened spin $s=1/2$ remains at the zero mode,  and 
a strong coupling regime 
shown in Fig.~\ref{fig:Doublet}~(c) where the perfect screening 
happens as formation of a singlet pair between the zero mode 
and the first orbital. 
When the weak coupling regime is realized, 
we should have a free electron behavior in the Dirac modes, 
which is not renormalized. 
While, in the strong coupling regime, we should find a magnetic 
moment of $1\mu_{B}$ 
mainly in the Dirac mode out of the zero mode. Here we assumed that 
the $g$ factor is 2 and $\mu_{B}$ is the Bohr magneton. 

The numerical result in the last section tells that 
the half-filling, {\it i.e.} $N_{first}=1$, is kept, even when $\mu$ 
is slightly shifted around the Dirac point, 
if stability of the screening happens owing to finite $\tilde{J}_{l=1}$. 
Therefore, the large $|J|$ area in the V-shaped region 
is the strong-coupling regime. 

At this stage, one might suspect that the present result 
is qualitatively different from those discussed in 
the pseudogap Kondo problem.\cite{0034-4885-76-3-032501} 
Often the problem is analyzed using 
the single-site impurity Anderson model (SIAM), 
where an impurity site is embedded in a pseudogapped semi-metal. 
Owing to the disappearing density of states at the 
charge neutral point, magnetic interaction mediated by 
the hybridization term between the impurity site and the 
pseudogapped metal disappears. 
The heuristic renormalization group approach  
concludes that an unscreened spin remains to keep the system 
with the doublet ground state. 
Therefore, the Kondo screening is concluded 
to be lost for the model at the half-filling. 

In our model, since the magnetic interaction is mediated by 
two-body scattering terms $V_{(0,k)\rightarrow (pq)}$, 
the interaction strength $\tilde{J}_{l=1}$ remains to be finite. 
The exchange is mediated by a kind of the super exchange process, 
where scattering of electrons via the Coulomb interaction 
causes virtual processes among unoccupied Dirac modes. 
We can explain the existence of the process as 
effective perturbation processes using a discretized spectrum 
as illustrated in Fig.~\ref{fig:Doublet}. 
On the contrary, the hybridization between the zero mode 
and the Dirac electrons is always zero in our modeling. 
Therefore, we call the mechanism ``scattering-driven Kondo screening''. 
This picture also tells us that the large cell limit is 
inevitably in the strong-coupling regime, as follows. 

Following the unitary transformation 
introduced in Sec.~\ref{sec:level2_2-2}, we have 
another third zero energy mode in 
resulted renormalized spectrum, $\tilde{\epsilon}_k$. 
(Fig.~\ref{fig:Doublet}~(b)) 
Owing to $J_{k,k'}$, we have a local singlet within two sites of TIAM, 
{\it i.e.} the zero mode and the first orbital, in the strong coupling regime. 
Then, an unpaired electron appears in another zero energy mode. 
(Fig.~\ref{fig:Doublet}~(c)) 
This crossover behavior from the weak to the strong coupling limit 
allows to have the doublet ground state in the whole parameter 
space considered. We may note that the real magnetic regime 
with possible higher spin states appears, only when $J_{kk'}$ behaves 
ferromagnetically. This can happen when $J^{direct}$ works. 

Interestingly, the super-exchange process works most effectively, when 
only a few Dirac modes around the Fermi level come close to 
the zero mode energetically. 
This condition is also met at the half filling. 

As noted, the hybridization term, $\tilde{H}_{hyb}$, is determined by $\epsilon_k$. 
When the system size grows, two low-lying 
Dirac states around the Dirac point approach to zero. 
In this limit, $V_{21}$ goes to zero. 
Then, by taking the large size limit, 
no matter small $\tilde{J}_{l=1}$ is, 
as far as $\tilde{J}_{l=1}$ is finite, 
we have a strong coupling regime 
for the large system size. 

The remaining Dirac electrons are, at the same time,  
renormalized in its spectrum, {\it i.e.} from $\epsilon_k$ to $\tilde{\epsilon}_k$. 
The number of the modes in $\tilde{\epsilon}_k$ is reduced by one 
from that of $\epsilon_k$. 
Then, even though the free electron picture of the Dirac cone 
gives an even-numbered spectrum of $\epsilon_k$, 
the renormalized Dirac band has another zero-energy mode decoupled 
from the zero mode in the strong correlation regime. 
Thus, we have a renormalized Dirac-cone state in TIAM, which is different from 
a semi-metallic state in an unscreened regime of SIAM. 

When $\mu$ is changed by the order of $\tilde{J}_{l=1}$, 
the half-filling nature of the extra localized mode, {\it i.e.} $N_{first}=1$, 
is lost as shown in Fig.~\ref{fig:Fig.1} (a). 
However, even when the magnetic screening disappears, 
the strong on-site interaction $U_0$ of the zero mode may keep the 
localized spin 1/2 at the center of $V_{111}$. (Fig.~\ref{fig:Fig.1} (b))
Therefore, around the Dirac point, 
we may tune the local magnetic property of $V_{111}$ 
only by changing the charge donation condition. 
As well-known, the gate bias can control $\mu$ rather easily 
for the graphene device structure. 

%%%%%%%%%%%%%%%%%%%%%%%%%%%%%%%%%%%%%%%%%%%%%%%%%%%%%

%%%%%%%%%%%%%%%%%%%%%%%%%%%%%%%%%%%%%%%%%%%%%%%%%%%%%
\section{CONCLUSIONS}
In conclusion, we derived a model for the zero mode 
of the hydrogenated atomic vacancy of graphene, $V_{111}$. 
The model is given by a Kondo Hamiltonian, where 
the two-particle scattering processes produce anti-ferromagnetic 
super exchange between the zero mode spin and 
low-lying modes of the Dirac cones. 
By way of a unitary transformation, 
we introduced a two-site impurity Anderson Hamiltonian, 
by which we traced the behavior of the model 
with continuous-time quantum Monte-Carlo calculation. 
The calculation results of the model suggests that 
the spin 1/2 on the zero mode on the $V_{111}$ structure may be magnetically 
screened by the Dirac electrons at low temperatures. 

Since the total system is a Kramers doublet at the half-filling, 
the remaining electron occupies the renormalized Dirac spectrum 
of odd numbered modes. In our many-body representation, 
however, the renormalized Dirac cones keeps its nature as 
a zero-gap semiconductor. 

Our result strongly suggests an interesting behavior of $V_{111}$, 
which had been created in a gate-biased graphene channel of electronic devices. 
The present simulation tells that the zero mode appears at the Dirac point, 
whose spectrum at around the $V_{111}$ structure can be controlled 
qualitatively by the filling control. 
Behavior of the spin 1/2 can be easily controlled by 
the external charge donation in the graphene variants. 

%%% Friedel oscillation
In this paper, we focused on $V_{111}$. In this structure, 
a single zero mode appears at the band center. 
When the electron (or hole) doping in the host graphene becomes 
large enough to create the Fermi surface structure, 
the electron-hole symmetry is broken. 
The Friedel oscillation in graphene\cite{PhysRevB.82.193405,PhysRevB.88.205416} 
would then be determined by a way reconstructing a self-consistent 
charge distribution, $n({\bf r})$. 
When a secondary order parameter 
in magnetism grows at the same time, our multi-reference 
density functional theory requires to determine the four current density 
as a source of the internal electro-magnetic field. 
A magnetic phase created by the RKKY interaction would, then, be 
determined in a self-consistent way. 

%%% LS
The description adopted in this paper takes several approximations {\it a priori}. 
This is because we looked at a many-body effect caused by 
``a pure zero mode'' of graphene. 
People have already derived enhanced spin-orbit coupling 
driven by the $\sigma$-$\pi$ mixing at a hydrogen impurity,\cite{PhysRevLett.103.026804,PhysRevLett.108.206808,PhysRevLett.110.246602}
whose signal in nature was reported.\cite{Balakrishnan2013}
When we introduce an effective single-particle Hamiltonian 
taking the spin-orbit coupling into account, the Kondo-type magnetic 
screening effect derived in this paper might explain a reason 
for the seemingly weak spin scattering by 
a localized electron in the zero mode. Furthermore, 
magnetic field direction dependence of the electron spin resonance 
and detection of photo induced quantum oscillation 
would complete our understanding of the impurity state experimentally. 

Our strategy of modeling is, however, hold for a wider class of 
defects of graphene, or pseudogapped Kondo systems. 
Analysis of graphene vacancy and atom-adsorbed graphene, 
as well as direct comparison with the experiments 
will be discussed elsewhere, by which nature of correlation effects 
in the zero mode of Dirac spectrum shall be further explored. 

%%%%%%%%%%%%%%%%%%%%%%%%%%%%%%%%%%%%%%%%%%%%%%%%%%%%%

%%%%%%%%%%%%%%%%%%%%%%%%%%%%%%%%%%%%%%%%%%%%%%%%%%%%%
\begin{acknowledgments}
The authors greatly thank stimulating discussions 
and fruitful comments by Dr. M. Ziatdinov, Dr. Y. Kudo, 
Prof. S. Fujii, Prof. M. Kiguchi, and Prof. T. Mori. 
They also appreciate helpful comments by Dr. H. Ueda. 
The calculations were partly done 
in the computer centers of Research Center for Information Technology, 
Kyushu University and 
Institute for Solid State Physics, University of Tokyo.
This work is supported by the Grand-in-Aid for scientific research 
(No. 235404080, No. 26107526, and No. 26400357) from MEXT, Japan. 
\end{acknowledgments}
%%%%%%%%%%%%%%%%%%%%%%%%%%%%%%%%%%%%%%%%%%%%%%%%%%%%%

\appendix
\section{Bandstructure results by DFT-LDA}
\label{Appendix_DFT_LDA}

\begin{figure}[htbp] 
\begin{center} 
\includegraphics[width=7cm]{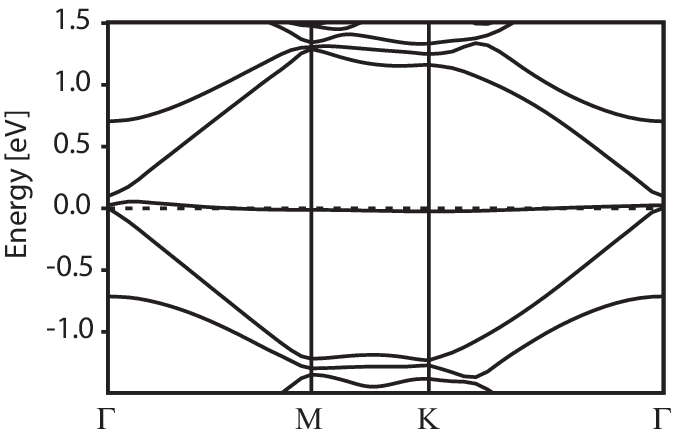} 
\end{center} 
\caption{The LDA band structure of $V_{111}$ in a $6\times6$ super cell.} 
\label{fig:Fig.LDA_band_a} 
\end{figure}

\begin{figure}[htbp] 
\begin{center} 
\includegraphics[width=7cm]{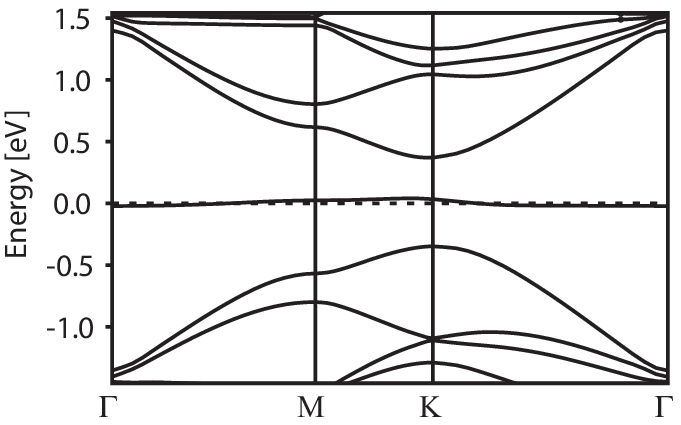} 
\end{center} 
\caption{The LDA band structure of $V_{111}$ in a $7\times7$ super cell.} 
\label{fig:Fig.LDA_band_b} 
\end{figure}

\begin{figure}[htbp] 
\begin{center} 
\includegraphics[width=7cm]{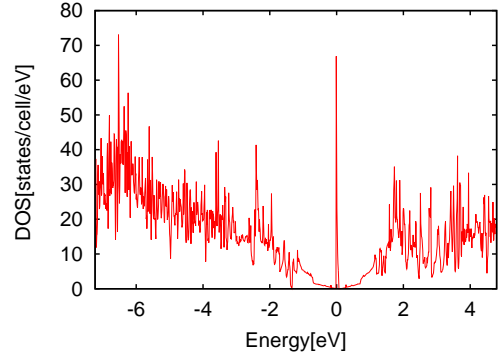} 
\end{center} 
\caption{The LDA density of states for $V_{111}$ in a $6\times6$ super cell.} 
\label{fig:Fig.LDA_DOS} 
\end{figure}

We show band structures of $V_{111}$ in two super cells 
given by the LDA calculations. The Born-von Karman boundary condition 
is adopted. 
Fig.~\ref{fig:Fig.LDA_band_a} is for the $6\times 6$ cell, and 
Fig.~\ref{fig:Fig.LDA_band_b} is for the $7\times 7$ cell. 
The density of states of the former structure is shown in 
Fig.~\ref{fig:Fig.LDA_DOS}. 

\begin{figure}[htbp] 
\begin{center} 
\includegraphics[width=6cm]{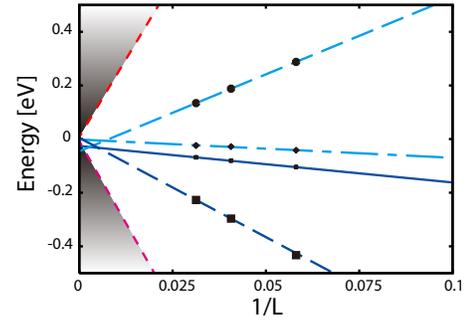} 
\end{center} 
\caption{Scaling relation of relevant energy for a $(3n+1)\times(3n+1)$ series of 
the supercells. The band bottom and 
the band top of the flat band (BBFB and BTFB) are given by the data 
connected by the solid line and the dot-dashed line. 
The shaded areas represent the energy range of the Dirac modes 
at the $\Gamma$ point in each super cell with the side-length $L$. 
The gap by the vacancy creation approaches to zero 
in the limit of $L\rightarrow \infty$. 
The dashed lines are the top (or bottom) of the Dirac branch appearing 
at the $K$ (or $K'$) points in FBZ of the super cell.} 
\label{fig:Fig.Gap_scale} 
\end{figure}

The size scaling of some relevant energies in the band structure is shown in 
Fig.~\ref{fig:Fig.Gap_scale}. Here, only $(3n+1)\times(3n+1)$ super cells are 
considered. In the limit of $L\rightarrow \infty$, 
the energy of the bottom of the flat band given by 
the blue line points an energy slightly below the gap closing point 
of the Dirac modes. By counting the electron correlation effect, however, 
our effective many-body model 
keeps the zero mode as a half-filled localized orbital. 

\section{A definition of a $\pi$ orbital on a deformed graphene}
\label{Appendix_pi_orbital} 

For the deformed graphene, we are able to define a $\pi$ network 
only by geometry of the atomic structure within a linear combination 
of atomic orbital  (LCAO) scheme. 

The direction of a $\sigma$ bond of a carbon atom 
is given by either local C-C or C-H direction. 
We have three direction vectors ${\bf a}_j^{i} = (a_{1j}^i,a_{2j}^i,a_{3j}^i)$ 
with $j=1,2,3$ for the $i$-th carbon atom. 
These vectors determine three $\sigma$-like orbitals, where 
each hybridized $\sigma$ wave function is given by 
\begin{equation}
\psi_{\sigma,j}^i({\bf r})
=C_{2s,j}^i \phi_{2s}^i({\bf r}) + \sum_{k=1,2,3} C_{2p,k,j}^i \phi_{2p,k}({\bf r}),
\end{equation}
where an index $k=1,2,3$ corresponding to $x$, $y$, $z$. 
Here, $\phi_{2s}^i({\bf r}) $ and $\phi_{2p,k}^i({\bf r}) $ are 
the local atomic orbital in a spatially fixed Cartesian coordinate. 
The coefficients are given by 
\begin{eqnarray}
\displaystyle C_{2s,j}^i &=& \frac{1}{C_j^i}
\sqrt{\frac{|{\bf a}_{j'}^{i}\cdot{\bf a}_j^{i}||{\bf a}_{j''}^{i}\cdot{\bf a}_j^{i}|}{|{\bf a}_{j'}^{i}{\bf a}_{j''}^{i}|}},\\
C_{2p,k,j} &=& \frac{1}{C_j^i}a_{kj}^i,\\
C_{j}^i&=&\sqrt{\frac{|{\bf a}_{j'}^{i}\cdot{\bf a}_j^{i}||{\bf a}_{j''}^{i}\cdot{\bf a}_j^{i}|}{|{\bf a}_{j'}^{i}{\bf a}_{j''}^{i}|}+\sum_{k=1,2,3}(a_{kj}^i)^2}.
\end{eqnarray}
Here $j$, $j'$, and $j''$ are mutually different. 
Note that, for $sp^2$-like hybridization, where 
the relative angle between two different direction vectors 
${\bf a}_{j}^i$ and ${\bf a}_{j'}^i$ are almost $2\pi/3$, 
the inner product ${\bf a}_{j}^i \cdot {\bf a}_{j'}^i$ is negative valued. 

Now, we have a unique local $\pi$ wave function, 
\begin{equation}
\psi_{\pi}^i({\bf r}) = D_{2s}^i\phi_{2s}^i({\bf r}) 
+\sum_{k=1,2,3}D_{2p,k}^i \phi_{2p,k}({\bf r}), 
\end{equation}
on each carbon atom, where $\psi_{\pi}^i({\bf r})$ is 
orthogonal to three $\psi_{\sigma,j}^i({\bf r})$. 
Thus, the coefficients, $(D_{2s}^i,D_{2p,k}^i)$, are determined 
uniquely except for a global phase factor, 
so that it is normalized and orthogonal to 
$(C_{2s,j}^i, C_{2p,k}^i)$ with $k=1,2,3$. 
We have also 1s orbitals $\phi_{1s}^l ({\bf r})$ for three 
hydrogen atoms ($l=1,2,3$). 

Using this definition of the $\pi$ orbital, 
we can introduce a topological mapping from the $t$ model 
to another tight-binding model in LCAO of 
$\sigma$ and $\pi$ orbitals. 
The dimension of a subspace given by $\{\psi_{\pi}^i\}$ is $N_C$, 
while that of LCAO is $4N_C+3$. 
Therefore, we have a submatrix of the LCAO Hamiltonian 
to be compared with $H_{TBM}$ in Eq.~(\ref{TBM0}). 
In a deformed structure, we have a finite transfer term  
from $\sigma$ to $\pi$ orbital, as well as 
a long range transfer term. 
The mapping from the second model to the $t$ model 
is defined as continuous reduction 
in both the $\sigma$-$\pi$ hybridization and 
the long-range $\pi$-$\pi$ transfer. 

\section{Mappings between zero modes for $3n \times 3n$ supercells} 
\label{Appendix_3n_mapping}

If one is required to consider another series of 
$3n \times 3n$, we may i) introduce an effective magnetic flux 
to change the gauge along an in-plane direction, 
ii) choose a $\Gamma$ point only calculation to have an effective band gap, 
and iii) apply evaluation in the same manner as discussed below. 
When we see a convergence of the obtained effective model 
for $n\rightarrow \infty$, the description in the thermodynamic 
limit is achieved. 

If one requires direct identification of the zero mode in 
$3n \times 3n$ super cells, there is another way, which is 
realized by incorporating two Wannier transformations. 
The zero mode, $\phi_0({\bf r})$, may be given by adopting the Wannier transformation 
of the center flat band. This is always possible, since there is an energy window 
required for the definition. (Fig.~\ref{fig:Fig.LDA_band_a})  
By the second Wannier transformation of the whole $\pi$ bands, 
we have defined $\pi$ orbitals $\phi_i({\bf r})$ for $V_{111}$. 
The set of $S_{\pi}= \{ \phi_i({\bf r}) \}$ is a complete set of the $\pi$ bands, 
by which the zero mode, $\phi_0({\bf r})$, is well expanded. 
There is a way to derive a non-orthogonal basis of $\{\phi_0,\{\phi_l\}\}$, 
where we may choose $(N_C-1)$ orbitals, $\phi_l({\bf r})$, 
from $S_{\pi}$. By applying an orthogonalization method, 
we can have an orthogonalized set of $S_{\pi}'=\{\phi_0,\{\tilde{\phi}_l\}\}$. 
Then, we have a unitary transformation connecting $S_{\pi}$ to $S_{\pi}'$. 
Utilizing the transformation, we can construct a Kondo Hamiltonian 
written in the basis of $S_{\pi}'$. 

\section{Super exchange mechanism for the zero mode} 
\label{Appendix_4n_superexchange}

In this appendix, we introduce an intuitive method to derive 
a super exchange process between the zero mode and 
a low-lying Dirac mode. 
For the purpose, considering the spectrum, 
$\varepsilon_k$ for the $k$-th Dirac mode, owing to the correlation effect, 
we assume that the electron occupation is one for 
the lowest $k=\pm1$ modes, 
and that the zero mode is singly occupied. 
The half electron occupation in the $k=1$ mode 
and a hole in the $k=-1$ mode can emerge naturally 
in a correlated state. 

We consider a state with a singlet pair, 
\begin{equation}
|\Psi_1\rangle = \frac{1}{\sqrt{2}}
(c^\dagger_{0\uparrow}c^\dagger_{k=1\downarrow}
-
c^\dagger_{0\downarrow}c^\dagger_{k=1\uparrow})
|\tilde{0}\rangle. 
\end{equation}
Here the state $|\tilde{0}\rangle$ denotes 
a filled Dirac sea for $k<-1$ with an up electron 
occupying the lowest $k=-1$ mode. 
\begin{equation}
|\tilde{0}\rangle = c^\dagger_{k=-1,\uparrow}\prod_{k<-1} 
c^\dagger_{k,\uparrow} c^\dagger_{k,\downarrow} |0\rangle. 
\end{equation} 
For the present purpose, 
we may select an up state of a doublet without loss of generality. 
In our description, the empty modes of $k>1$ and 
the filled modes of $k<-1$, which are 
empty hole states, contribute to produce 
intermediate virtual states in the super exchange processes. 

\begin{figure}[htbp] 
\begin{center} 
\includegraphics[width=6cm]{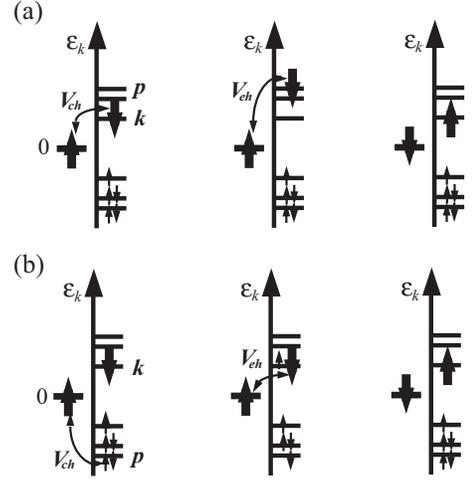} 
\end{center} 
\caption{Super-exchange processes  
in (a) an electron channel and (b) a hole channel 
for the zero mode $k=0$. The electron in a mode $k$ 
interacts with the localized electron at the zero mode 
via the correlated hopping $V_{ch}$ to make 
a virtual state. The exchange hopping $V_{eh}$ 
exchanges two spins, which completes the super exchange. } 
\label{fig:Fig.Super_proc} 
\end{figure}

Since the effective one-body Hamiltonian is diagonalized, 
the virtual state is created by the effective two-body Hamiltonian, 
which appears as the quantum fluctuation. 
Since the zero mode with large $U_0$ is in the strong correlation regime, 
the occupation of the zero mode 
does hardly fluctuate even in the virtual state. 
Therefore, a relevant process happens via 
the correlated hopping $\hat{H}_{(0k)\rightarrow(0p)}^{ch}$ 
and the exchange hopping $\hat{H}_{(0k)\rightarrow(p0)}^{eh}$. 
We consider an effective Hamiltonian for these processes, 
\begin{eqnarray}
\lefteqn{\hat{H}_{(0k)\rightarrow(0p)}^{ch} } \nonumber \\
&=& V_{(0k)\rightarrow(0p)}\left(
c^\dagger_{0,\uparrow} c^\dagger_{p,\downarrow}
c_{k,\downarrow} c_{0,\uparrow} 
+c^\dagger_{p,\uparrow} c^\dagger_{0,\downarrow}
c_{0,\downarrow} c_{k,\uparrow} 
\right), \nonumber \\
\lefteqn{\hat{H}_{(0k)\rightarrow(p0)}^{eh} } \nonumber \\
&=& V_{(0k)\rightarrow(p0)}\left(
c^\dagger_{p,\uparrow} c^\dagger_{0,\downarrow}
c_{k,\downarrow} c_{0,\uparrow} 
+c^\dagger_{0,\uparrow} c^\dagger_{p,\downarrow}
c_{0,\downarrow} c_{k,\uparrow} 
\right). \nonumber 
\end{eqnarray}
When $p>k$, 
the action of the effective two-particle Hamiltonian is, 
\begin{eqnarray}
\lefteqn{ 
\left( \hat{H}_{(01)\rightarrow(0p)}^{ch} 
+\hat{H}_{(01)\rightarrow(p0)}^{eh}\right)
|\Psi_1\rangle } \nonumber \\ 
&=&
\left( V_{(0k)\rightarrow(0p)} + V_{(0k)\rightarrow(p0)} \right)
\frac{1}{\sqrt{2}}
(c^\dagger_{0\uparrow}c^\dagger_{p\downarrow}
-
c^\dagger_{0\downarrow}c^\dagger_{p\uparrow})
|\tilde{0}\rangle . \nonumber 
\end{eqnarray}
As an intermediate state owing to the quantum fluctuation, 
a singlet is formed between the singly-occupied $p$ mode 
and the zero mode. 
For comparison, let us also test a triplet state. 
\begin{equation}
|\Psi_2\rangle = \frac{1}{\sqrt{2}}
(c^\dagger_{0\uparrow}c^\dagger_{k=1\downarrow}
+
c^\dagger_{0\downarrow}c^\dagger_{k=1\uparrow})
|\tilde{0}\rangle,
\end{equation}
which behaves as, 
\begin{eqnarray}
\lefteqn{ 
\left( \hat{H}_{(01)\rightarrow(0p)}^{ch} 
+\hat{H}_{(01)\rightarrow(p0)}^{eh}\right)
|\Psi_2\rangle } \nonumber \\ 
&=&
\left( V_{(0k)\rightarrow(0p)} - V_{(0k)\rightarrow(p0)} \right)
\frac{1}{\sqrt{2}}
(c^\dagger_{0\uparrow}c^\dagger_{p\downarrow}
+
c^\dagger_{0\downarrow}c^\dagger_{p\uparrow})
|\tilde{0}\rangle . \nonumber 
\end{eqnarray}
These are processes by quasi-electron excitation. 
(Fig.~\ref{fig:Fig.Super_proc} (a))
There are other processes by quasi-hole excitation, 
where a filled $k=1$ state and a hole at a mode with $p<-1$ appear. 
(Fig.~\ref{fig:Fig.Super_proc} (b))

Using these results, we can readily obtain the expression 
of the super exchange energy mediated by 
the intermediate states with $\epsilon_p$ 
as an energy difference between two states, 
$|\Psi_1\rangle$ and $|\Psi_2\rangle$. 
\begin{eqnarray}
\label{J_super_energy}
\lefteqn{\Delta E(\Psi_2)-\Delta E(\Psi_1)}
\nonumber \\
&=&\sum_{|\varepsilon_p|>\varepsilon_{k=1}}
\frac{2( V_{(0k)\rightarrow(0p)} V_{(0p)\rightarrow(k0)}
+V_{(0k)\rightarrow(p0)} V_{(p0)\rightarrow(k0)} )}
{\varepsilon_p-\varepsilon_{k=1}}. \nonumber 
\end{eqnarray}
The final form of the super exchange interaction 
between the zero mode and a $k$ mode has the next form 
as a scattering of the electron at the $k$ mode to 
another $k'$ mode. 
\begin{equation}
\hat{H}_{kk'}^{super} 
= J_{kk'}^{super} 
(c^{\dag}_{0\sigma} \vec{\sigma}_{\sigma \sigma '} c_{0\sigma '})
\cdot (c^{\dag}_{k\tau} \vec{\sigma}_{\tau \tau '} c_{k'\tau '}),
\end{equation}
with
\begin{eqnarray}
\label{J_super_exp}
J^{super}_{kk'}
&=&\frac{1}{2}\sum_{|\epsilon_p|>\epsilon_c}\left\{
\frac{V_{(0k)\rightarrow(0p)}V_{(0p)\rightarrow(k'0)}}
{\varepsilon_p-\varepsilon_k} \right. \nonumber \\
&&+\left. \frac{V_{(0k)\rightarrow(p0)}V_{(p0)\rightarrow(k'0)}}
{\varepsilon_p-\varepsilon_k} \right\}.
\end{eqnarray}
Here $\epsilon_c$ is a cut off to separate 
the low-energy phase space active for the exchange scattering 
and higher energy virtual states. 
In a finite size system, we may choose only the 
lowest excitation of $k=\pm 1$ for the low-energy modes. 
While we see quasi-degeneracy at the $k=1$ (or $k=-1$) mode 
often in a super cell result. Therefore, we keep an expression 
with the cutoff energy. 

Relevant information on the estimation of 
$J^{super}_{kk'}$ is the next. 
One might suspect that the exchange hopping amplitudes, 
$V_{(0k)\rightarrow(p0)}$, might not be enough large to have 
a finite value of $J^{super}_{kk'}$. 
We have tested overlap of the wave function of 
the $t$ model. 
Let's introduce eigen states and 
eigen functions of the $t$ model. 
In this appendix, we omit the spin index in each expression 
for simplicity. 
\begin{eqnarray}
\hat{H}_{TBM} |\psi_k\rangle 
&=& \tilde{\varepsilon}_{k} |\psi_k\rangle, \\
|\psi_k\rangle &=& \sum_{i}\psi_{k,i} C^\dagger_{i} |0\rangle, \\
\psi_k({\bf r}) &=& \sum_{i}\psi_{k,i} \phi_i({\bf r}),  \\
\int d^3r \phi_i^*({\bf r})\phi_j({\bf r}) &=& \delta_{ij}.
\end{eqnarray}
The last equality is the 
orthogonality among the Wannier orbitals. 
A next integral is estimated. 
\begin{eqnarray}
\lefteqn{
\tilde{V}_{(0k)\rightarrow (p0)}
}\nonumber \\
&=& \frac{e^2}{2}
\int \! \! \! \int d^3r d^3r' 
\frac{\psi_p^*({\bf r})\psi_0^*({\bf r}')\psi_k({\bf r}')\psi_0({\bf r})}{|{\bf r}-{\bf r}'|}
\nonumber \\
&=& 
\frac{e^2}{2}\sum_{i_1,i_2,i_3,i_4}
\psi_{p,i_1}^*\psi_{0,i_2}^*\psi_{k,i_3}\psi_{0,i_4}
\tilde{v}_{i_1,i_2,i_3,i_4},
\\
\lefteqn{
\tilde{v}_{i_1,i_2,i_3,i_4}
}\nonumber \\
&=&
\int \! \! \! \int d^3r d^3r' 
\frac{\phi_{i_1}^*({\bf r})\phi_{i_2}^*({\bf r}')\phi_{i_3}({\bf r}')\phi_{i_4}({\bf r})}
{|{\bf r}-{\bf r}'|}.
\end{eqnarray}
Using an approximate expression 
$\tilde{v}_{i_1,i_2,i_3,i_4}=
\tilde{v}(|{\bf R}_{i_1}-{\bf R}_{i_2}|)\delta_{i_1,i_4}\delta_{i_2,i_3}$ 
with the position of the $i$-th carbon atom, ${\bf R}_{i}$, and 
\[\tilde{v}(|{\bf R}_{i_1}-{\bf R}_{i_2}|)
=\left\{ \begin{array}{ll} v_0 & {\rm for}\; {\bf R}_{i_1} = {\bf R}_{i_2} \\ 
\frac{v_0}{|{\bf R}_{i_1}-{\bf R}_{i_2}|} & 
{\rm for}\; {\bf R}_{i_1} \neq {\bf R}_{i_2} \end{array}
\right.,\]
we may evaluate asymptotic behavior of 
$\tilde{V}_{(0k)\rightarrow (p0)}$ as well as 
$\tilde{V}_{(0k)\rightarrow (0p)}$ for a large super cell. 
In this estimation, however, we cannot determine 
the absolute value of the integral, or $v_0$, since 
$\phi_{i}({\bf r})$ is not explicitly given. 
Therefore, we choose a finite system to adjust $J^{super}$ 
in this evaluation to the value by the DFT result. 
The result is shown in Fig.~\ref{fig:Fig_Interaction_scale} of the main text. 

\begin{figure}[htbp] 
\begin{center} 
\includegraphics[width=7cm]{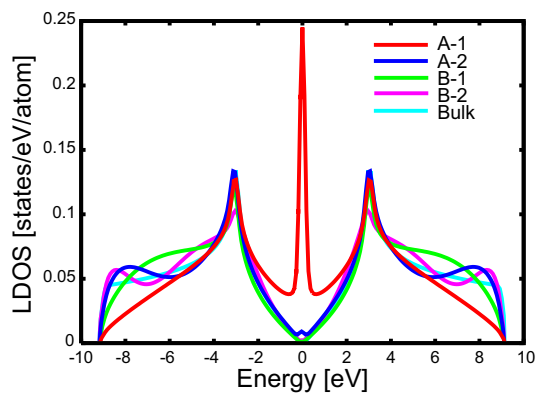} 
\end{center} 
\caption{The local density of states (LDOS) of 
the $V_{111}$ structure. The value is estimated by 
the $t$ model with the $100\times 100$ super cell. 
Each lines depict LDOS at the first carbon site in the A sub-lattice 
($A-1$), the nearest neighbor B-site ($B-1$), 
the second nearest A-site ($A-2$), the third nearest B-site ($B-2$), 
and a site apart from the vacancy (Bulk). } 
\label{fig:Fig.LDOS} 
\end{figure}

Using the estimation by the $t$ model, 
we can derive another key. Let us show the local density of states 
around $V_{111}$. (Fig.~\ref{fig:Fig.LDOS}) 
In the $A$-sites neighboring to the vacancy, we have 
a clear indication of the zero mode as a sharp peak of LDOS 
at the center of the spectrum. We see a clear enhancement 
of LDOS at the $A-1$ site, even when the energy is rather in the Dirac spectrum. 
We have also a slight signal of enhancement at the $A-2$ site. 
This result suggests us that local overlap between a low-lying Dirac electron 
and the zero mode electron around $V_{111}$. Owing to this overlap 
of the wave function, we have non-negligible contribution 
$\tilde{V}_{(0k)\rightarrow (p0)}$ as well as 
$\tilde{V}_{(0k)\rightarrow (0p)}$. 

\bibliographystyle{unsrt}
\bibliography{bib_paper1}

\end{document}